\documentclass[onecolumn]{svjour2}
 
\usepackage{graphicx}
\usepackage{endnotes}
\usepackage{authblk}                                           
\usepackage{amssymb}
\usepackage{bbm}                                               
\usepackage{tensor}                                            
\usepackage{amsmath}
\usepackage[usenames,dvipsnames]{color}
\usepackage{setspace}                           
\usepackage{hyperref}
\usepackage{epigraph}
\usepackage[utf8x]{inputenc}
\usepackage[left= 1in,right=1in,top=1in,bottom=1in]{geometry}                                  

\hypersetup{
    colorlinks=true,         
    linkcolor=blue,          
    citecolor=red,        
    urlcolor=Violet             
}

\let\oldmarginpar\marginpar
\renewcommand\marginpar[1]{\oldmarginpar{\color{red}\raggedright\scriptsize #1}}
\newcommand{\pb}[2]{\ensuremath{\lf\{#1,#2 \rt\}}}
\newcommand{\mean}[1]{\ensuremath{\lf\langle #1 \rt\rangle }}
\newcommand{\diby}[2]{\ensuremath{\frac{\partial #1}{\partial #2}}}
\newcommand{\ddiby}[2]{\ensuremath{\frac{\delta #1}{\delta #2}}}
\newcommand{\tind}[3]{\ensuremath{t^{\phantom{#1}#2}_{#1#3}}}
\newcommand{\Tind}[3]{\ensuremath{T^{\phantom{#1}#2}_{#1#3}}}

\newcommand{\ket}[1]{\ensuremath{\lf| #1 \rt> }}

\def\lf {\ensuremath{\left}}
\def\rt {\ensuremath{\right}}
\def\mA {\ensuremath{\mathcal A}}

\def\lie {\ensuremath{\mathcal L}}
\def\ham {\ensuremath{\mathcal H}}

\title{Symmetry and Evolution in Quantum Gravity}

\author{Sean Gryb \and Karim Th\'ebaault}

\institute{Sean Gryb
	    \at Institute for Theoretical Physics, Utrecht University, Leuvenlaan 4, Utrecht, The Netherlands
		\and
	    Sean Gryb
	     \at Radboud University Nijmegen, Institute for Mathematics, Astrophysics and Particle Physics, \\\email{s.gryb@hef.ru.nl}
		\and
	    Karim Th\'ebault
		\at MCMP, Ludwig Maximilians Universit\"{a}t, Ludwigstrasse 31, D-80539, Munich, Germany, \\\email{karim.thebault@gmail.com}
		}
\begin{document}

\maketitle

\begin{abstract}
    We propose an operator constraint equation for the wavefunction of the Universe that admits genuine evolution. While the corresponding classical theory is equivalent to the canonical decomposition of General Relativity, the quantum theory contains an evolution equation distinct from standard Wheeler--DeWitt cosmology. Furthermore, the local symmetry principle --- and corresponding observables --- of the theory have a direct interpretation in terms of a conventional gauge theory, where the gauge symmetry group is that of spatial conformal diffeomorphisms (that preserve the spatial volume of the Universe). The global evolution is in terms of an arbitrary parameter that serves only as an unobservable label for successive states of the Universe. Our proposal follows unambiguously from a suggestion of York whereby the independently specifiable initial data in the action principle of General Relativity is given by a conformal geometry and the spatial average of the York time on the spacelike hypersurfaces that bound the variation. Remarkably, such a variational principle uniquely selects the form of the constraints of the theory so that we can establish a precise notion of both symmetry and evolution in quantum gravity.
\end{abstract}
 
\section{Introduction}

It could be taken to be a basic requirement of any faithful canonical quantization procedure that the behaviour which characterized the classical system may be recovered from the quantum theory in the appropriate semi--classical limit. Notions of symmetry and evolution, in particular, are expected to be coherent across the classical and quantum formalisms. The case of gravity, however, does not easily allow us to satisfy this seemingly basic requirement. As has been discussed over the last fifty years \cite{Bergmann:1959,DeWitt:1967,Isham:1992,and2010}, when canonical quantization is applied to general relativity (after it has been put into its Hamiltonian form \cite{Dirac:1958b,ADM:1960}) the formalism produced fails to implement an analogue of the classical notions of either evolution or symmetry. In the classical formalism both dynamical evolution and foliation symmetry are implemented via the Hamiltonian constraint functions, $\mathcal{H}(x)$ (where $x$ is a point on a spatial hypersurface). Yet, the standard Dirac procedure implies that in the quantum formalism $\mathcal{H}(x)$ appears solely as an operator annihilating the wave function: this then implies that both physical Hilbert space states and observable operators do not evolve. There is thus no manifest sense in which a correlate to classical evolution is implemented in such a timeless Wheeler--DeWitt formalism for quantum gravity. The situation with regard to symmetries is similarly troubling --- as noted by Moncrief \cite[p.2962]{moncrief:1990} --- `what, after all, is a quantum spacetime and does such an object admit a representation in terms of different space-like slicings?' On its own, the Wheeler--DeWitt formalism offers us little hope of answering such four dimensional symmetry questions since it provides us only with a wavefunction over a space of spatial metrics. Our criteria for a faithful quantization procedure are thus not fulfilled; such a na\"{i}ve formalism for quantum gravity does not provide us with a satisfactory notion of either symmetry or evolution. 

Several different avenues can be pursued in the face of this difficulty.\footnote{The list of references given here is not intended to be comprehensive. Readers are encouraged to consult the classic review \cite{Isham:1992} and a modern update \cite{and2010} for a more detailed and wide--ranging catalogue of approaches to the \textit{Problem of Time}. For our purposes, we feel that it is best to treat the `Problem of Time' as an entirely separate issue to the `Problem of the Arrow of Time'. However, for an interesting exploration of a possible connections between the two problems see \cite{kiefer:1995,kiefer:2007,kiefer:2012}.} One option is to adapt the notion of quantization in a way that respects spacetime diffeomorphism invariance. Examples of research programs that follow this option include the causal set approach \cite{bombelli:1987,dowker:2005,henson:2006},  topos theory \cite{isham:2000,isham:2002}  and the causaloid approach \cite{hardy:2007}. Another option is to try to redefine the notion of observables at the classical level and apply something closer to traditional canonical quantization techniques. Notable examples of this strategy are the complete and partial observables scheme \cite{Rovelli:1990,Rovelli:1991,Rovelli:2002,Dittrich:2006,Dittrich:2007} and the master constraint programme \cite{DittrichThie:2006,Thiemann:2006,Thiemann:2007}. A third option, pursued for example by \cite{Husain:2012PRL,Giesel:2012DustLQG,BrownKuchar:1994Dust,IshamKuchar:1984Dust}, is to add an additional ``dust'' field that can serve as an internal clock for the system. More conservatively, rather than including additional fields, one might look for a formulation of General Relativity that clearly identifies the dynamical degrees of freedom so that standard quantization methods can be employed. Such a decomposition is pursued by the causal dynamical triangulation approach \cite{loll:2001,ambjorn:2001}, for example, which relies on a discretization of the geometry. Here, we will present a formulation that does not rely on a discretization but still provides an unambiguous identification of the degrees of freedom. 

Our proposal follows from a suggestion of York \cite{York:GR_boundary} whereby the independently specifiable data in the action principle of General Relativity is given by a conformal geometry and the spatial average of the York time on the spacelike hypersurfaces that bound the variation. Remarkably, such a variational principle uniquely selects the form of the constraints of the theory so that we can establish a precise notion of both symmetry and evolution in quantum gravity. In the context of our formalism, the requirement for a faithful quantization can be expressed in terms of the preservation of the \textit{characteristic behaviour} of York's ontology within the quantum formalism. This can be achieved via the implementation of a \textit{Relational Quantization} methodology which was first introduced in \cite{gryb:2011}. For the case of gravity, when understood in York's terms, Relational Quantization leads explicitly to an operator constraint equation for the wavefunction of the Universe that admits genuine evolution. The global evolution is explicitly characterized in terms of an arbitrary parameter which serves only as an unobservable label for successive states of the Universe. Furthermore, the local symmetry principle --- and corresponding observables --- of the \textit{quantum} theory have a direct interpretation in terms of the conventional gauge theoretic relationship, where the corresponding \textit{classical} gauge symmetry group is that of spatial conformal diffeomorphisms (that preserve the spatial volume of the Universe).\footnote{The technical steps for achieving this were largely developed in \cite{barbour_el_al:physical_dof,gryb:shape_dyn}.} Significantly, while in our approach the classical theory is equivalent to the canonical decomposition of General Relativity, the quantum theory makes predictions that are distinct from Wheeler--DeWitt cosmology. In particular, for homogeneous cosmology, both the conformal factor \emph{and} the scalar field are to be promoted to operators in the quantum theory. This is in contrast to standard Wheeler--DeWitt cosmology where the homogeneous Wheeler--DeWitt equation (i.e., the quantum Friedmann equation) is deparametrized with respect to one of these. Such a formal difference will have empirical consequences in the early universe.

On a formal level, our procedure is nearly identical to the proposal made in \cite{brown:gr_time}, which led to the development of unimodular gravity \cite{Henneaux_Teit:unimodular_grav,Unruh:unimodular_grav,Unruh_Wald:unimodular,Smolin:unimodular_grav} as a solution to the problem of time. This programme sparked some interest but was met with the criticism \cite{Kuchar:unimodular_grav_critique} that it necessarily selects an unphysical preferred foliation. Our proposal can be seen as a revival of this approach with a more precise motivation stemming from a careful analysis of the quantization procedure for reparametrization invariant theories. However, our proposal goes further because, as we will see, the insistence on identifying York's ontology in the variational principle of General Relativity leads directly to a preferred foliation. As such, our method favours a clean notion of symmetry over manifest spacetime invariance. From the point of view of General Relativity, it is the new conformal symmetry principle we obtain that selects the preferred foliation required by our approach. It should be stressed however that, although our approach is nearly identical to that used in the toy model presented in \cite{brown:gr_time}, the way this method is implemented in gravity differs importantly from the unimodular approach both in the treatment of the local symmetry and the choice of the evolution parameter. For the complete details, see Section~\ref{sec:Rel GR}.

We note that our approach is closely connected to the following intuition regarding the danger as to misidentification of the classical observables,  again from Moncrief\footnote{Moncrief's statement is closely related to the analysis of the classical Hamiltonian constraints given in   \cite{Barbour:timelessness,Barbour:2008,Pons:2010,pitts:2013}} \cite[p.2962]{moncrief:1990}:
\begin{quote}
One often hears of the desirability of constructing a complete set of ``observables" for the gravitational field, i.e., a maximal independent set of functions of the ADM canonical variables $\{g_{ab},\pi^{ab}\}$ which Poisson commute with all of the constraints...However, such observables, by themselves, merely provide an unambiguous set of labels for the space-times in question and do not yield, without further information, the geometrical properties of these space-times. In this respect they are somewhat analogous to the complete sets of initial positions and momenta that label the solutions of any problem in Hamiltonian mechanics.
\end{quote} 
and, furthermore, to the suspicion of Kucha\u{r} \cite[p.178]{kuchar:prob_of_time} (see also \cite{Kuchar:1991,kuchar:time_int_qu_gr}), that timelessness is not an inevitable consequence of the canonical quantum gravity formalism: 
\begin{quote}
The problem of time smacks of an eleatic paradox...[T]he intrinsic
metric and the extrinsic curvature are genuinely different on
different instants of an Einstein spacetime. Quantum theory does not
prohibit change either: an ordinary Schr\"{o}diner equation tells us
how change occurs. So the people who tried to quantize geometry surely
made some slip... 
\end{quote}
According to our analysis, the `slip' concerns a misidentification of the classical degrees of freedom of gravity so that a the Wheeler--DeWitt quantization is not, in our terminology, a faithful quantization. In terms of the quantum theory, the important feature that our proposal captures, which is not captured by the standard Wheeler--DeWitt equation, is that no particular choice of clock must be made amongst classical observables. This way, all classical observables can be promoted to operators in the quantum theory (forcing a particular classical observable to be a clock forces it to have a definite, i.e., non-quantum, value after quantization). The combination of York's ontology for gravity and relational quantization then leads to a reformulation of certain facets of the problem of time such that they admit direct solution. The following section will present the general argument of this paper in more detail.   

\section{General Argument}\label{sec:gen argument}

\subsection{Kinds of Degrees of Freedom}

In this section, we develop a classification scheme for symmetries. To be precise, we must first establish our notation and clarify what we mean by a classical theory. For this reason, we will review standard methodology for describing physical systems using variational principles.

Consider a classical theory consisting of a \emph{configuration space} $\mA$ upon which we define a \emph{Lagrangian}, $\mathcal L$, and its associated \emph{action}, $S[\gamma]$ --- which is the integral of the Lagrangian along some \emph{path}, $\gamma$, in $\mA$. We then define a variational principle that extremizes $S[\gamma]$ given some \emph{boundary} or \emph{initial conditions} that we must specify for the problem. The classical history of the system, $\gamma_\text{cl}$, is given by the extremal path. These mathematical structures can be taken to correspond to a physical system only once an interpretation has been given to the elements of the configuration space \textit{and} one has established a correspondence between the physical behaviour that characterises the system and the formulation of the variational problem in terms of explicit boundary or initial conditions.    We will elaborate upon the significance of the conditions on the variational problem in a moment. Classical physics requires these minimal ingredients, in some combination or another. We collect them below:
\begin{enumerate}
    \item The appropriate mathematical structures:
    \begin{itemize}
	\item A configuration space $\mA$.
	\item A Lagrangian on $\mA$.
	\item A trial curve $\gamma$ in $\mA$ with some parametrization $t$.
	\item Boundary conditions for the variation of the action $S[\gamma] = \int_\gamma \mathcal L$.
	\item The existence of a unique $\gamma_{\text{cl}}$, compatible with the variational conditions, such that the variation $\lf. \delta S[\gamma] \rt|_{\gamma_\text{cl}} = 0$.
    \end{itemize}
    \item An interpretation of:
    \begin{itemize}
	\item the elements of $\mA$.
	\item the boundary conditions for $\delta S[\gamma]$.
    \end{itemize}
\end{enumerate}

Once one has specified these ingredients, one can construct additional structures that are not formally required to model a classical system,    but are of great value both practically and conceptually.\footnote{Of course, one could alternatively formulate one's methodology solely in terms of these new structures but we will not do so here.} We can define a dual map $\star$ through the Legendre transform so that the cotangent bundle, or phase space, $T^\star_q \mA = \Gamma(q,p)$ is such that $q \in \mA$ and
\begin{equation}
    p = \ddiby{S}{\dot q},
\end{equation}
where $\dot q \in T_q \mA$.

The phase space, $\Gamma$, can be equipped with a Poisson bracket $\pb {q_i} {p_j} = \delta_{ij}$, where $i$ and $j$  label either discrete or continuous indices. Using this bracket, functions on phase space, $f: \Gamma \to \mathbbm{R}$, induce Hamilton vector fields, $v_f \in T\Gamma$, via the definition $v_f(\cdot) = \pb \cdot f$. If the Legendre transform is invertible, then $\dot q$ can be expressed uniquely in terms of phase space quantities and the Hamiltonian $H = \dot q \cdot p - L(q,p)$ can be defined. The flow of the Hamilton vector field of the Hamiltonian can be used to reconstruct $\gamma_\text{cl}$ using the inverse of the Legendre transform. 

We have been careful and general with our definitions so that we may now be able to speak of general properties of symmetries. A symmetry has its origin in an additional complexity that can arise in the methodology outlined above. Invariably it is associated with \textit{some} form of redundancy occurring in the relationship between our mathematical formalism and the system to which it corresponds. More precisely, symmetries occur when the formal combination of configuration space and variational principle do not uniquely represent the \textit{characteristic behaviour} of the system itself, as embodied by the physical degrees of freedom and physical boundary conditions. Thus, under our account, the interpretation of symmetries must be dictated by the comparison between formal redundancies, on the one hand, and the behaviour of the physical degrees of freedom, on the other. It is the manner in which the redundancies occur that allows us to classify the type of symmetry that is present.  Explicitly, we may use one criteria relevant to the degrees of freedom and one criteria relevant to the variational principle in order to construct a classification scheme with five possible outcomes. Our scheme will provide a framework for the classification of symmetries that both recovers the important implications of existing distinctions and provides new insights. In particular, our classification scheme will be able to accommodate the existence of \emph{hidden} symmetries within the formalism. The existence of such symmetries was first observed in \cite{gryb:shape_dyn}, where spatial conformal symmetry was identified as a hidden symmetry of General Relativity. 

To be more explicit, we will restrict our discussion to two classes of symmetries: i) those that have a Lie group action on $\mA$, and ii) those that generate reparametrizations of curves on $\mA$. Although, this may seem somewhat restrictive, we will see, in Sec~\ref{sec:Gravity}, that General Relativity can be expressed as a theory with symmetries of this kind. Remarkably, since Yang--Mills gauge theories can also be represented this way, the whole of the standard model coupled to gravity (plus many interesting systems in mechanical and condensed matter systems) are treatable by our considerations. We thus believe that our classification scheme covers a very general set of physically motivated theories containing symmetry, including the main case of interest in our paper: gravity.

The first criterion in our classification scheme pertains to the particular independent function on phase space associated with a given degree of freedom. If the action \textit{does not} change when $\gamma$ is varied with respect to this function,\footnote{Because we are speaking about the properties of the action evaluated along curves on $\mA$, which are insensitive to the internal structure of individual points on $\mA$, these symmetries correspond to traditional \emph{global} symmetries.} then there is an orbit on phase space along which the action is invariant. This orbit is generated by the flow of the Hamilton vector field of the phase space function that is canonically conjugate to that degree of freedom. This situation automatically implies a symmetry, and we will call a symmetry of this kind \emph{manifest}. There are two further possibilities: either there is no symmetry at all, or there is a \textit{hidden symmetry}. This third possibility will be discussed below (see Case~4). As is well-known, and as we will see explicitly in Section~\ref{sec:Noether}, in general the generator of a symmetry is always a constant of motion. Its value is determined by the details of the variational problem. 

The second criterion relates to the type of variation used to define the variational principle. If the variational principle is one in which no conditions are imposed on the degree of freedom we say it is a \textit{free variation} (for that degree of freedom). From a technical point of view, this will mean imposing that the action be \emph{completely} independent of the degree of freedom in question. As explained in section~\ref{sec:Classical Gauge Theory}, at the classical level, this can be achieved by requiring that the momentum of that degree of freedom vanish (on top of the standard Euler--Lagrange relations).\footnote{Note that this is a \emph{stronger} requirement then requiring that the variation of that degree of freedom on the endpoints of the variation is free, as suggested in \cite{barbour_el_al:physical_dof}.} These conditions should be enforced when the degree of freedom in question has no physical interpretation in the system. That is, the degree of freedom was introduced into the theory for mathematical or conceptual convenience, and, thus, there is no physical input from the characteristic behaviour of the system to constrain the relevant variable's value. There also exists an alternative type of variational principle that is such that it \textit{does} put definite restrictions on the degree of freedom in question. These restrictions are derived directly from the characteristic behaviour of the system. We will call these kinds of variations \emph{fixed} and will describe how to implement them in explicit examples in Section~\ref{sec:Noether}. Our two criteria can now be used to classify degrees of freedom and symmetries into the following five cases:\footnote{This list is not intend to be exhaustive, but rather cover all cases of physical interest.} 

\textbf{Case 1}: If the degree of freedom corresponds to a manifest symmetry and the variation is free, then the degree of freedom is \emph{gauge} and the symmetry is a \emph{gauge symmetry}. These symmetries occur in Electromagnetism, Yang--Mills theories and the Standard Model.\footnote{In Section~\ref{sec:Gravity}, we illustrate how General Relativity can be cast into a gauge theory of this kind where the symmetry group is that of the volume preserving conformal diffeomorphisms.} A gauge degree of freedom is a product of redundancy in the formalism, and has usually been explicitly introduced into a physical theory  solely for mathematical convenience. Although it may play a more profound role in physical theory than is presently clear in the context of quantum theory, classically at least a gauge variable is inherently otiose. That the generator of a gauge symmetry should be a constraint makes intuitive sense since the corresponding variable value cannot be fixed through any possible experimental input. Further examples of systems that exhibit symmetries of this kind and illustrate the properties just described are given in Section~\ref{sec:Noether}.

\textbf{Case 2}: Degrees of freedom that correspond to a manifest symmetry but a fixed variation lead to conserved charges and we will refer to them as \emph{conservation symmetries}. A simple example of conservation symmetries are the ones treated by Noether's First Theorem \cite{Fatibene:hole_argument}. The main difference between a gauge symmetry and a conservation symmetry is that the degree of freedom associated with these symmetries has a physical interpretation. Each symmetry has an associated constant of motion, which is the generator of the symmetry. This constant is fixed by the conditions on the variation. It has a definite interpretation in terms of the characteristic behaviour of the system. Examples of conservation symmetries that exhibit the properties described here are given in Section~\ref{sec:Noether type}.

\textbf{Case 3}: The simplest case is if there is no manifest symmetry and the variation is fixed. This case corresponds to a conventional dynamical degree of freedom. Its initial or boundary conditions are specified by the variational principle and it evolves according to the flow of the Hamiltonian.

\noindent The most non-trivial case is if there is no manifest symmetry and the variation is free. This situation splits into at least two distinct cases.

\textbf{Case 4}: If there is no manifest symmetry, the variation is free, \textit{but} there exists an additional manifest symmetry in the theory that is  second class with respect to the symmetry in question (i.e., the Poisson brackets between the constraints which generate the manifest symmetry and those which generating the hidden symmetry is not weakly zero), then there is a \emph{hidden} symmetry in the system. If this is the case, the Hamiltonian can be modified (without changing the physical predictions of the theory) in such a way that the first symmetry becomes manifest. This is called \emph{symmetry trading} and has been used to construct the \textit{Shape Dynamics} formalism introduced in \cite{gryb:shape_dyn}. The general theory of symmetry trading is developed in \cite{Gomes:linking_paper}. We will see in detail how this plays out in General Relativity in Section~\ref{sec:Gravity}. The existence of hidden symmetries that can be traded for other manifest symmetries in a particular theory is a relatively recent observation in the literature. The classification scheme given here is, in part, advantageous over standard textbook classifications of symmetries because of how it highlights this possibility.

\textbf{Case 5}: If there is no other symmetry in the theory or if the symmetry is not second class with respect to another manifest symmetry, then the situation is more complicated and the system is very likely to be inconsistent. 

We collect all five cases in Figure~\ref{fig:table of dof}:

\begin{figure}[h]
    \begin{center}
\begin{tabular}{|c|c|c|l|}\hline
\textbf{Case} & \textbf{Variation} & \textbf{Symmetry} & \textbf{Classification}\\\hline\hline
1 & Free & Manifest & Gauge Symmetry\\\hline
2 & Fixed & Manifest & Conservation Symmetry\\\hline
3 & Fixed & None & Conventional Degree of Freedom\\\hline
4 & Free & Hidden & Tradeable Symmetry\\\hline
5 & Free & None   & Possible Inconsistency\\\hline
    \end{tabular}
    \end{center}
    \caption{Classification of different degrees of freedom.}\label{fig:table of dof}
\end{figure}

It is important to note the following features of our definition of a gauge symmetry:
\begin{itemize}
\renewcommand{\labelitemi}{$-$}
    \item First, our definition principally relies  on the nature of the variational principle; namely, that a gauge degree of freedom must be varied freely. This follows from the fact that a gauge degree of freedom should be distinguished solely by whether or not its has a correspondence to some aspect of the characteristic behaviour of a physical system. That this is determined by the nature of the variation is a crucial observation due to Barbour and is based on an earlier observation of Poincar\'e \cite{poincare:principle}. Details can be found in \cite{barbourbertotti:mach,barbour:bm_review} or, more recently, in \cite{sg:dirac_algebra}. Furthermore, our definition implies that the treatment of a degree of freedom as physical or unphysical depends on how the mathematical framework is to be used to model the physical system in question. In some situations, the degree of freedom can be treated as gauge, while in others, it may correspond to a physical background field.
    \item Second, our definition gives us a more accurate and better motivated method for classification than the `textbook definition' of gauge symmetries --- i.e., that a gauge symmetry is a \emph{local} symmetry of an action. Rather, we will see in Section~\ref{sec:Classical Gauge Theory} that the fact that gauge symmetries are \emph{local} symmetries of the action is a consequence of the free variation.
    \item Third, our definition does not mention the first class or second class nature of the constraints. This is because second class constraints can either be eliminated using Dirac's procedure \cite{dirac:lectures,henneaux_teit:quant_gauge} or treated as a particular gauge fixing of a first class system. The latter can be achieved either in a manifest way or according to a \emph{gauge-unfixing} procedure \cite{Vytheeswaran:gauge_unfixing,henneaux_teit:quant_gauge}.
    \item Fourth, and most significantly for what follows, our definition of a gauge symmetry does \emph{not} just follow from the non-invertibility of the Legendre transform (although it implies it). Instead, the definition relies on the physical interpretation of the degrees of freedom and their variation. As we will see below, the Legendre transform can fail to be invertible even when all the degrees of freedom in $\mA$ are physical. In this situation, invocation of the notion of `gauge freedom' is inappropriate and misleading.\footnote{A related argument leading to the same conclusion is given in \cite{Barbour:2008}.}
\end{itemize}

\subsection{Reparametrization Invariance}

A theory is said to be \emph{reparametrization invariant} when the action is independent of the choice of parametrization of $\gamma$. A simple argument (which we give in Section~\ref{sec:Rep Class}) shows that, for such theories, the Hamiltonian is proportional to a constraint and will always vanish along $\gamma_\text{cl}$. The presence of constraints is implied by the failure to invert the Legendre transform. Physically, this is because the magnitude of $\dot q$ depends upon the choice of $t$, which is irrelevant to the physics. This means that the momenta cannot uniquely determine the velocities, resulting in constraints. However, (unless there are other constraints) there is still every reason to believe that \emph{all} the coordinates of $\mA$ are physical in the sense that their value can be read off, at some instant, from a device in the system. This is true no matter which coordinate of $\mA$ is used to parametrize the evolution. This leads us to the following key observation: \emph{reparametrization symmetries do not imply the existence of gauge degrees of freedom in our description of a physical  system.} Indeed, in reparameterization invariant theories, the boundary variation is of the \emph{fixed} type because all configuration space coordinates correspond to degrees of freedom that can be given a direct  interpretation in terms of the characteristics of the system. Thus, reparametrization invariant theories fall into the second category listed in Figure~\ref{fig:table of dof}. Because they are manifest symmetries with fixed variation, they are grouped more naturally with conservation symmetries than with gauge symmetries. The differences between these two kinds of symmetry are subtle but important in quantization. This will be the main theme of Section~\ref{sec:Noether}.

Let us attempt to further justify this significant claim made above. What is unphysical in reparametrization invariant theories is the parametrization of $\gamma$: saying that a theory is reparametrization invariant means precisely that the parametrization of paths found within our formalism is not related to a corresponding feature within the physical system being described. Formally this equates to the magnitude of the velocities being unphysical. This is precisely what we want since we are dealing with a system where the characteristic behaviour is such that duration is derived through change. Given this, one might suggest that the length of the velocities should be taken as `gauge'. However, this would imply an odd number of degrees of freedom on phase space, which would leave us with a \textit{non-canonical} formalism where the physical phase space (i.e., that space parametrized by the `true' degrees of freedom) fails to be a symplectic manifold. How to deal with the mathematical structure of such a theory --- in particular with regard to quantization --- is entirely unclear.

Instead, we suggest that it is better to think of the total energy as a constant of motion, like a conserved charge --- which does not propagate because it is constant ---but which is also \textit{not} a gauge degree of freedom. Then, the Hamiltonian constraint should be interpreted as an identity satisfied by the classical evolution but should not be imposed formally as a constraint. Consequently, we \emph{should not} directly apply the Dirac procedure when quantizing constraints that generate global reparametrizations.

It is important to note that this is will not change anything significant with regard to the classical theory. The classical evolution will still live on surfaces of constant energy. These surfaces will be parametrized by initial or boundary conditions that specify the total energy of the system and we will model precisely the same physical systems. However, the quantum theory will undergo drastic changes since superpositions of energy eigenstates are allowed.

In a previous paper \cite{gryb:2011}, we presented arguments towards a `Relational Quantization' methodology which implemented this non-standard interpretation of reparametrization. One purpose of this paper is to provide a more concrete and mathematically well grounded justification for Relational Quantization. In particular, in Section~\ref{sec:Formal Constructions}, we will draw upon the general results of Section~\ref{sec:Noether}, to describe the treatment of reparametrization invariant theories using Relational Quantization. Then, in Section~\ref{sec:Gravity}, we will show how our approach can be made compatible with gravity. 

Before this, and to conclude this general introductory section, we will motivate a particular identification of the characteristic behaviour of General Relativity. That our Relational Quantization procedure will lead one towards a consistent understanding of symmetry and evolution within quantum gravity will follow unambiguously from this characterization.

\subsection{York's Ontology: Physical Degrees of Freedom for Gravity}\label{sec:York Ontology}

The procedure described above for classifying degrees of freedoms relies on a careful consideration of the variational principle used in the theory to make definite predictions. This, in turn, relies heavily on the physical interpretation of the configuration space, $\mA$, and our understanding of the characteristic behaviour of systems described by the theory. In order to be able to apply our procedure to gravity, we need to be able to specify precisely what the physical degrees of freedom and physical boundary conditions are. This translates into asking what degrees of freedom are independently specifiable on the boundary of the variation or, rather, ``What is fixed on the boundary in the action principles of General Relativity?'' This question was posed by Wheeler to York, and is addressed in \cite{York:GR_boundary}. In our paper, we will adopt the view taken in Section~4 of that paper: what is fixed on the boundary is the conformal spatial geometry and the mean of the York time (which is proportional to the trace of the extrinsic curvature of $\Sigma$). We will refer to this identification of the independent degrees of freedom of gravity as \emph{York's ontology}.

In \cite{York:GR_boundary}, York performs a decomposition of the spatial metric and its conjugate momentum into pieces that correspond to separating out the conformal part of the spatial geometry from the other parts. In terms of the momentum density, $\pi^{ab}$, conjugate to the spatial metric $g_{ab}$, this decomposition is inspired by the spin decomposition, introduced in York's seminal work \cite{York:cotton_tensor,York:york_method_long,York:york_method_prl}, of $\pi^{ab}$ into a pure trace piece, $\pi^{ab}_\text{tr}$; a pure longitudinal piece, $\pi^{ab}_\text{long}$; and a transverse--traceless piece, $\pi^{ab}_\text{TT}$, such that
\begin{equation}
    \pi^{ab} = \pi^{ab}_\text{tr} + \pi^{ab}_\text{long} + \pi^{ab}_\text{TT}.
\end{equation}
York notes that the spatial diffeomorphism constraints of General Relativity can be solved by requiring $\pi^{ab}_\text{long}$ to vanish, while the Hamiltonian constraint itself can be treated as an equation for the local scale factor of $g_{ab}$. As was appreciated first in \cite{barbour_el_al:physical_dof} (and more explicitly in \cite{gryb:shape_dyn}), the form of the Hamiltonian constraint used to solve for the conformal factor in York's method, is actually an equation for the part of the conformal factor that \emph{does not} change the volume (in the spatially compact case) and, resultantly, is a gauge-fixing for a constraint that requires $\pi^{ab}_\text{tr}$ to be proportional to a constant, $K$. York then identifies this constant (which is conjugate to the spatial volume), along with the conformal geometry (which is conjugate to $\pi^{ab}_\text{TT}$) as the independently specifiable data in the variational principle of General Relativity.

To our knowledge, this is currently the only known decomposition of the degrees of freedom of General Relativity that can be used to (provably) solve the initial value problem given an \emph{arbitrary} initial metric and momentum. Since this is precisely what we need to be able to apply our classifications scheme for degrees of freedom and symmetries, York's proposal is the only method we know of that can be used for our analysis. In addition to this practical motivation, we note that York's decomposition has found important applications in many gravitational systems, particularly when these systems lack global symmetries. It is the most widely used tool, for example, for solving the generic initial value problem in numerical relativity \cite{cook:living_rev_num_rel}. Furthermore, it is a key step in the identification of gauge-independent variables in perturbative cosmology \cite{Mukhanov:cosmology_review}. Finally, the decomposition also singles out the `shape dynamics' ontology advocated by Barbour \cite{JuliansReview} on the basis of Mach's principles. In Section~\ref{sec:Gravity}, we will show that York's ontology leads to a precise notion of symmetry and evolution in quantum gravity illuminating its utility as a tool for studying non-trivial gravitational systems. Before proceeding with this analysis however, we will motivate our treatment of symmetries by giving a precise definition of `free' and `fixed' variations, illustrated through simple examples. On a first reading, the technical details can be skipped by passing to the summary sections \ref{sec:Noether Summary} and \ref{sec:gauge summary}.

\section{Manifest Symmetries in Finite Dimensions} \label{sec:Noether}
\subsection{Best Matching Procedure}

The main purpose of this section is to examine the difference between \emph{free} and \emph{fixed} variations in the presence of a manifest symmetry. The case where the symmetry is \textit{not} manifest, but hidden (in the sense defined above), will be treated in the context of GR in Section~\ref{sec:Gravity}. Also, we will make a short comment at the end of this section on how the results presented here are expected to generalize to known gauge field theories. As stated earlier, we will restrict ourselves to symmetries realized by a Lie group action $\mathcal G$ on $\mA$. For symmetries of this kind, \emph{fixed} variations lead to identities expressing the presence of conserved charges while \emph{free} variations lead to linear Gauss-like constraints that generate the symmetry. To be completely explicit, we will consider only finite dimensional systems. In Section~\ref{sec:Rep Class}, we will go on to show how reparametrization invariant theories should be treated in analogy to the conservation symmetry case and contrast this with the standard Dirac approach (applicable to gauge theories). The material in this section is a much more thorough and precise account of some of the material presented in Section~2 of \cite{sg:dirac_algebra}.

To study our model, we will use the formalism of best matching,\footnote{Specifically, we are referring to the part of the best--matching procedure involved with spatial relationalism. For details on best matching see \cite{barbourbertotti:mach,barbour:scale_inv_particles} or \cite{sg:dirac_algebra}.} although a variety of other techniques could be used. The reason for utilsing best matching is that it allows for an explicit parametrization of the symmetric coordinate (and its conjugate momentum) via a phase space extension. This will prove to be unavoidable for the consistent treatment of reparametrization invariant theories. Thus, best matching provides a useful tool for treating this case. We will develop the required aspects of best matching as they are needed so that no previous knowledge of best matching will be required to follow the argument.

We consider a finite dimensional configuration space, $\mA$, with coordinates, $q^a$ on which a Lie group  $\mathcal G$ acts such that
\begin{equation}\label{eq:G transform}
    q^a \to G(\theta^\alpha)^a_b q^b
\end{equation}
is a global invariance of $S[\gamma]$. Here, $G^a_b$ is a some $\dim (\mA)$ representation of $\mathcal{G}$ parametrized by $\theta^\alpha \in T\mathcal G = \mathfrak g$, an element of the Lie algebra of $\mathcal G$ in some chart. The indices $a,b$ range from 1 to $\dim (\mA)$ and $\alpha$ ranges from 1 to $\dim (\mathfrak g)$. By global invariance, we mean that $S[\gamma]$ is invariant under time-\emph{independent} choices of $\theta^\alpha$. There are several ways to express this in terms of an explicit mathematical condition on the action. We will follow the procedure laid out by best matching, which involves first artificially parameterizing the symmetry by introducing an auxiliary field.

To do this, we extend the configuration space $\mA(q^a) \to \mA_\text{e} (q^a, \theta^\alpha)$ to include the group parameters $\theta^\alpha$ and insert the transformation \eqref{eq:G transform} into the Lagrangian $ L( q^a, \dot q^a) \to L(G^a_b q^b, G^a_b \dot q^b + \dot G^a_b q^b)$. Calling $G^{-1a}_{\phantom{-1}b}$ the right-inverse of $G^a_b$ --- defined by $G^{-1a}_{\phantom{-1}c} G^c_b = \delta^a_c$ --- we make the convenient definition
\begin{equation}
    \Tind \alpha a b (\theta^\alpha) := G^{-1a}_{\phantom{-1}c} \diby{G^c_b}{\theta^\alpha},
\end{equation}
which, in general, depends upon $\theta^\alpha$. When evaluated about the group identity, $\theta^\alpha = 0$, this is equal to the generator, $\tind \alpha a b$, of the Lie algebra, $\mathfrak{g}$, of $\mathcal G$:
\begin{equation}
    \tind \alpha a b \equiv \lf. \Tind \alpha a b (\theta^\alpha) \rt|_{\theta^\alpha = 0}
\end{equation}
in a dim($\mA$) representation. Using this, we can write the transformed Lagrangian in terms of the covariant derivative\footnote{If $\mA_\text{e}$ is thought of as a fiber bundle, where the $\theta^\alpha$-directions are fibers over $\mA$, then $D$ defines a section of this bundle. We will see that, upon phase space reduction with free variation, this reduces to a covariant derivative on $\mA$, as in Yang--Mills gauge theory.} 
\begin{equation}
    D_\theta q^a = \dot q^a + \dot\theta^\alpha \Tind \alpha a b q^b.
\end{equation}
Since $G^a_b \dot q^b + \dot G^a_b q^b = G^a_b D_\theta q^b$, we have $ L(q^a,\dot q^a) \to L(G^a_b q^b, G^a_b D_\theta q^b)$. However, global gauge invariance implies
\begin{equation}
    \diby{L(G^a_b q^b, G^a_b \dot q^b)}{\theta^\alpha} = 0
\end{equation}
because $\dot\theta = 0$. In order for this to be true for all $\theta$, we must have
\begin{equation}
    L(q^a, \dot q^a) \to L(G^a_b q^b, G^a_b D_\theta q^b) = L(q^a, D_\theta q^b).
\end{equation}
Thus, the transformed Lagrangian on $\mA_\text{e}$ is obtained, in the case of a manifest symmetry, by replacing time derivatives of $q^a$ with covariant derivatives. 

The Legendre transform of the system can be performed by defining the momenta
\begin{equation}\label{eq:p def}
    p_a = \diby{L}{\dot q^a}
\end{equation}
and
\begin{equation}\label{eq:pi def}
    \pi_\alpha = \diby L {\dot \theta^\alpha} = \diby L {\dot q^a} \Tind \alpha a b q^b.
\end{equation}
This leads to an extension of the phase space $\Gamma(q,p) \to \Gamma_\text{e}(q,\theta; p, \pi)$ and the additional Poisson brackets
\begin{equation}
    \pb {\theta^\alpha} {\pi_\beta} = \delta^\alpha_\beta.
\end{equation}
The above formulas immediately imply the primary constraint
\begin{equation}\label{eq:kretch}
    C_\alpha \equiv \pi_\alpha - p_a \Tind \alpha a b (\theta)q^b \approx 0.
\end{equation}
The meaning of these constraints becomes clear when considering the flow generated by these constraints on $\Gamma_\text{e}$ in terms of an infinitesimal smearing $\epsilon^\alpha$:
\begin{align}\label{eq:inf trans}
    \delta_{\epsilon^\alpha} q^a &= \epsilon^\alpha \Tind \alpha a b q^b & \delta_{\epsilon^\alpha} p_a & = -\epsilon^\alpha p_b \Tind \alpha b a & \delta_{\epsilon^\alpha} \theta^\alpha &= -\epsilon^\alpha,
\end{align}
where $\delta_{\epsilon^\alpha} \cdot = -\epsilon^\alpha \pb{\cdot}{ \pi_\alpha - p_a \Tind \alpha a b (\theta)q^b}$ (the flow of $\pi_\alpha$ is unenlightening). Clearly, these are just the infinitesimal symmetry transformations $q^a \to G^a_b(\epsilon) q^b$ and $p_a \to G^{-1b}_{\phantom{-1}a}(\epsilon) p_a$ plus simple translations of $\theta^\alpha$. The sign is such that the two contributions exactly cancel to leave the transformed quantities $G^a_b(\theta)q^b$ and $G^{-1b}_{\phantom{-1}a}(\theta) p_b$ invariant. Thus, $\theta^\alpha$ can be interpreted as a compensator field for the symmetry.

The Hamiltonian is obtained from
\begin{equation}\label{eq:Ham1}
    H(q,\theta;p,\pi) = \dot q^a p_a + \dot \theta^\alpha \pi_\alpha - L(q^a,\theta^\alpha; p_a, \pi_\alpha)
\end{equation}
after eliminating $\dot q^a$ and $\dot \theta^\alpha$ in terms of $p_a$ and $\pi_\alpha$ using \eqref{eq:p def} and \eqref{eq:pi def}. We can use the symmetry properties of $L$ to show that $L(q^a, D_\theta q^a)$ is \emph{only} a function of $q^a$ and $p_a$. This is a result of the fact that $\diby{D_\theta q^a}{q^b} = \delta^a_b$ so that
\begin{equation}
    p_a = \diby{L}{\dot q^a} = \diby{L}{D_\theta q^b} \diby{D_\theta q^b }{\dot q^a} = \diby{L}{D_\theta q^a}.
\end{equation}
One can then solve for $D_\theta q^a$ as if it where $\dot q^a$ everywhere and make the appropriate substitutions. Then, we can write the total Hamiltonian as the sum
\begin{equation}
    H_\text{tot} = H + N^\alpha C_\alpha,
\end{equation}
where $N^\alpha$ is a Lagrange multiplier. Using a redefinition, $N^\alpha \to N^\alpha + \dot \theta^\alpha$, we can write $H$ in a way that doesn't depend on the phase space extension by absorbing all the terms in \eqref{eq:Ham1} that depend on $\theta^\alpha$ and $\pi_\alpha$ into the definition of $N^\alpha$ such that
\begin{equation}
    H (q,\theta;p,\pi) \to H(q,p) = \dot q^a p_a - L(q,p).
\end{equation}
This important property of the Hamiltonian is valid only because of the global symmetry of the action.

The system of Hamiltonian, $H$, and constraints, $C_\alpha$, can be seen to be first class by noting that it can be obtained by a canonical transformation of the manifestly first class system formed by the Hamiltonian and the (now) trivial constraint
\begin{equation}
    \pi_\alpha \approx 0.
\end{equation}
The canonical transformation $\Gamma_\text{e}(q,\theta; p, \pi) \to \Gamma_\text{e}(\bar q,\bar \theta; \bar p, \bar \pi)$ that achieves this is obtained from the type-2 generating functional
\begin{equation}\label{eq:finite can transf}
    F(q^a, \theta^\alpha; \bar p_a, \bar \pi^\alpha) = \bar p_a G^{-1a}_{\phantom{-1}b}(\theta) q^b + \theta^\alpha \bar \pi_\alpha,
\end{equation}
which performs the transform
\begin{align}
    \bar q^a &= G^a_b q^b & \bar p_a &= p_b G^{-1b}_{\phantom{-1}a} \\
    \bar\theta^\alpha &= \theta^\alpha & \bar\pi_\alpha &= \pi_\alpha - p_a \Tind \alpha a b q^b
\end{align}
leading to the desired result. Because we have a first class system, the Dirac constraint algorithm is complete.

The ability to define this type of canonical transformation suggests a `short-cut' for applying our procedure that starts directly from phase space. This short cut involves, first, extending the phase space by introducing compensator fields for the symmetry in question, then finding the constraints that restrict this degree of freedom. In the case of a symmetry generated by a Lie group (as above), this involves performing a canonical transformation that explicitly parametrizes that symmetry.

From this persecutive, the importance of our classification lies in helping identify the type of manifest symmetry --- fixed or free --- that we are dealing with. It is this categorization that determines how we should treat the extended Hamiltonian formalism and, in turn, the path towards the quantum theory. The following two subsections are devoted to consideration of the fixed and free cases, respectively.  

\subsection{Fixed Variations and Conservation Symmetries}\label{sec:Noether type}

The virtue of best matching, in this context, is that we gain explicit access to key structural features of the manifest symmetry. By performing the phase space extension, we gain access to the parts of the original phase space associated with the symmetry in question. It is then straightforward to manipulate these structures in any way that we like. In particular, we can explicitly define what we mean by \emph{fixed} variation. This formalism will make it clear how Noether's first theorem arises in the classical theory and how the Ward identities arise in the quantum theory. It will also suggest a methodology for dealing with reparametrization invariant theories.

\subsubsection{Classical Theory: Conserved Charges}\label{sec:classical fixed}

Classically, we require a fixed variation for all the variables of $\mA$. This means that all $q^a$'s should evolve according to the usual Euler--Lagrange equations. In the Hamiltonian picture, this means that we just require standard Hamiltonian evolution on the original phase space $\Gamma$. This can be achieved in our extended phase space picture by treating the auxiliary fields $\theta^\alpha$ and $\pi_\alpha$ as standard gauge degrees of freedom constrained by the first class constraints $C_\alpha \approx 0$ and by performing a phase space reduction. Because of the global symmetry of the action, solving the constraints will leave a mark on the reduced theory. This will be an expression of Noether's first theorem.

Before performing the phase space reduction, we identify the Dirac observables of the extended theory by recalling that, because of the infinitesimal action of the $C_\alpha$'s \eqref{eq:inf trans}, the $\theta^\alpha$'s act like compensator fields for the symmetry. Thus, the transformed functions
\begin{equation}
    f(G^a_b(\theta) q^b, G^{-1b}_{\phantom{-1}b}(\theta) p_b)
\end{equation}
will commute with $C_\alpha$ and, therefore, will be a basis for the Dirac observables of the theory.

The first step in the phase space reduction is to perform the gauge fixing
\begin{equation}
    \theta^\alpha = 0.
\end{equation}
This is a valid gauge fixing of $C_\alpha = 0$ because
\begin{equation}
    \pb {\theta^\alpha}{N^\beta C_\beta} = N^\alpha,
\end{equation}
which can be made to vanish by the \emph{unique} choice of smearing
\begin{equation}
    N^\alpha = 0.
\end{equation}
In this gauge, $\pi_\alpha$ is a conserved charge since\footnote{In other gauges, the explicit expression for the conserved charged can be obtained locally on phase space by isolating the canonical variable conjugate to the constraint $C^\alpha = 0$ using local Darboux coordinates. The different expressions for the conserved charge in different gauges are related by the canonical transformation generated by \eqref{eq:finite can transf}. They represent non-standard representations of the original ontology of the theory.}
\begin{equation}
    \dot \pi_\alpha = \pb{\pi_\alpha}{H_\text{tot}} = \pb{\pi_\alpha}{H(q,p)} = 0.
\end{equation}

Using the Dirac bracket, we can perform the phase space reduction by applying the constraints and gauge fixings strongly. For this gauge fixing, the Dirac bracket is trivial to compute because $\pb {\theta^\alpha}{ C_\beta} = \delta^\alpha_\beta$ is diagonal and, thus, easy to invert. Its action
\begin{equation}
    \pb \cdot \cdot_\text{Db} = \pb \cdot \cdot - \pb{\cdot}{\theta^\alpha} \pb{C_\alpha}\cdot + \text{permutations}
    \end{equation} 
is clearly equal to the action of the Poisson bracket on the original phase space $\Gamma(q,p)$. Thus, we can apply the gauge fixing $\theta^\alpha = 0$ \emph{strongly} and use the standard Poisson bracket on $\Gamma$. Now the constrains $C_\alpha$ take the simple form
\begin{equation}
    C_\alpha = \pi_\alpha - p_a \lf. \Tind \alpha a b (\theta)\rt|_{\theta = 0} q^b = \pi_\alpha - p_a \tind \alpha a b q^b.
\end{equation}
We complete the phase space reduction by treating the $C_\alpha = 0$ as a strong equation for $\pi_\alpha$. Because $\pi_\alpha$ does not appear anywhere else in the theory, this is inconsequential other than the fact that, as we just proved, they are conserved charges $\pi_\alpha = \pi_\alpha^0$. Thus, $C_\alpha = 0$ gives
\begin{equation}\label{eq:noether}
    \pi_\alpha^0 = p_a \tind \alpha a b q^b,
\end{equation}
which implies that $p_a \tind \alpha a b q^b$ are also conserved charges. This is Noether's first theorem. To see that these are really the conserved charges we are used to, it suffices to consider a couple concrete examples.

Consider the case where $\mA$ represents $n$ particles in $d$ dimensions. Then, the index $a$ can be split into $a = iI$, where $i= 1\hdots d$ ranges over spatial indices and $I=1\hdots n$ ranges over particle indices. The generator of translations on the ambient $\mathbbm R^d$, is just a spatial gradient acting separately on each particle. There are $d$ generators so that $\alpha = 1\hdots d$. If we call the coordinates that parametrize this symmetry $a^i$ and their conserved charges $\mathcal P_i$, then the constraints \eqref{eq:noether} take the form
\begin{equation}
    \mathcal P_i - \sum_{I=1}^n p_i^I \approx 0,
\end{equation}
where $p_i^I = p_{iI}$ is the momentum of the $I^\text{th}$ particle. For rotations, one can do something similar. In $d=3$ the generators take the cross product between $q_I$ and $p_I$. If we call the rotational coordinates $\phi^i$ and their momenta $\varphi_i$, then \eqref{eq:noether} become 
\begin{equation}
    \varphi_i - \sum_{I=1}^n \epsilon\indices{_i^j_k} p_i^I q^k_I  \approx 0,
\end{equation}
where $q^k_I = q^{ik}$ is the coordinate of the $I^\text{th}$ particle. Thus, for theories with manifest Euclidean symmetries, conservation of total linear and angular momentum are implied, as expected. The general expression \eqref{eq:noether} reflects the presence of general conserved charges.

Finally, we note that, after the phase space reduction, the Dirac observables $f(G^a_b q^b, G^{-1b}_{\phantom{-1}a} p_b)$ become simple functions, $f(q^a,p_a)$, the original phase space $\Gamma$, as expected.

\subsubsection{Quantum Theory}

After doing this preliminary work, many of the features of the quantum theory can be readily obtained. This is because the phase space extension and constraints expose many of the important formal structures implied by the manifest symmetry. To see this, we can simply perform the Dirac constraint quantization of the extended theory \cite{dirac:lectures}. Section~\ref{sec:Dirty Quant} describes how these methods can be applied to the specific context of the quantization of reparametrization invariant theories. 

Here we should note that despite its great heuristic value, the Dirac methodology for constraint quantization suffers from a number of formal defects. For instance, if the relevant constraint operator has a continuous spectrum there are strictly no zero eigenvectors, and so one cannot define the physical Hilbert space via the Dirac constraint equation. Such issues necessitate application of more sophisticated quantization techniques, such as a group averaging methodology. See \cite{isham:1984a,isham:1984b,henneaux_teit:quant_gauge,Giulini:1999a,Giulini:1999b,Thiemann:2007} for more details. While important, these issues are strictly tangential to the focus of the current paper: the innovation with regard to quantization concerns the classical starting point and the classification of constraints, and not the precise methodology for enacting quantization. Thus, we will here always talk in terms of Dirac quantization but with the implict proviso that a more rigorous methodology could be substituted provided the classification of degrees of freedom and constraints remains the same. 

The Dirac quantization of the extended theory can be achieved by promoting phase space functions to operators (using a particular operator ordering convention) and Poisson brackets to commutators. Thus,
\begin{align}
    \pb{q^a}{p_b} &= \delta^a_b \to [\hat q^a, \hat p_b]\Psi = i \hbar\delta^a_b \Psi \\
    \pb{\theta^\alpha}{\pi_\beta} &= \delta^\alpha_\beta \to [\hat \theta^\alpha, \hat \pi_\beta] \Psi = i \hbar \delta^\alpha_\beta \Psi.
\end{align}
These operators act on elements of the auxiliary Hilbert space $\mathcal H$. The Dirac prescription requires that the wavefunction, $\Psi$, belong to the \emph{physical} Hilbert space, $\mathcal{H}_\text{phys}$, satisfying the operator constraints $\hat C_\alpha \Psi = 0$, or
\begin{equation}\label{eq:Ward}
    \hat P_\alpha \Psi = \hat \pi_\alpha \Psi,
\end{equation}
where $\hat P_\alpha = \hat p_a \Tind \alpha a b (\hat \theta) \hat q_b$. In the Schr\"odinger picture, time evolution is generated by the Hamiltonian
\begin{equation}\label{eq:noether evolution}
    \hat H(\hat q, \hat p) \Psi = i\hbar \diby{}t \Psi.
\end{equation}
To be explicit, we will pick a configuration basis for the operators so that $\hat \theta^\alpha = \theta^\alpha$ and $\hat \pi_\alpha = -i\hbar \diby{}{\theta^\alpha}$ and the wavefunction can be treated as a function of $q^a$, $\theta^\alpha$, and $t$. We will choose the operator ordering
\begin{equation}\label{eq:P ordering}
    \hat P_\alpha = -i\hbar q^b \Tind \alpha a b \diby{}{q^a}
\end{equation}
for the generalized momentum operator $\hat P_\alpha$. 

If the spectrum of $\hat P_\alpha$ is known, the independence of the Hamiltonian on $\theta^\alpha$ suggests a separation ansatz
\begin{equation}
    \Psi(q^a,\theta^\alpha,t) = \sum_n c_n e^{i\theta^\alpha \pi_\alpha^n} \psi_n(q^a,t),
\end{equation}
which solves the quantum constraints \eqref{eq:Ward} provided $\psi_n(q^a,t)$ is an eigenvalue of $\hat P_\alpha$ with eigenvalue $\pi_\alpha^n$. If $\{\psi_n\} \, \forall\, n$ is a complete orthonormal basis for the spectrum of $\hat P_\alpha$, then one can define the inner product on $\mathcal H_\text{phys}$ as
\begin{equation}
    (\psi_n,\psi_m) = \delta_{nm}.
\end{equation}

We now note an important feature of the theory that will allow us to reduce the quantum theory from $\mathcal H \to \mathcal H_\text{phys}$. In terms of our ansatz, \eqref{eq:Ward} takes the form
\begin{equation}
    -i\hbar q^b \Tind \alpha a b (\theta) \diby{}{q^a} \psi_n(q^a,t) =  \pi^n_\alpha \psi_n(q^a,t)
\end{equation}
for all $n$ and the evolution equation \eqref{eq:noether evolution} becomes
\begin{equation}
    \sum_n e^{\theta^\alpha \pi_\alpha^n} \hat H \psi_n = \sum_n e^{\theta^\alpha \pi_\alpha^n} (i\hbar) \diby{\psi_n}{t}.
\end{equation}
The value of $\theta^\alpha$ does not affect the solutions of these equations, which are given in terms of $\psi_n(q^a,t)$ only. We can, thus, pick any gauge that we like for $\theta^\alpha$. Choosing $\theta^\alpha = 0$, we find that the eigenvalue equation reduces to
\begin{equation}
    -i\hbar q^b \tind \alpha a b \diby{}{q^a} \psi_n(q^a,t) =  \pi^n_\alpha \psi_n(q^a,t),
\end{equation}
which is the direct analogue of the Ward identity for the associated symmetry (assuming that no anomalies are present). The general solution of \eqref{eq:noether evolution} is then given by the superposition
\begin{equation}\label{eq:Psi decomp}
    \Psi(q^a,t) = \sum_n c_n \psi_n(q^a,t).
\end{equation}
The reduction $\mathcal H \to \mathcal H_\text{phys}$ in this basis is expressed by $\Psi(q^a,\theta^\alpha,t) \to \Psi(q^a,t)$ through the gauge choice $\theta^\alpha = 0$.

Observables are just given by self-adjoint operators acting on $\Psi(q^a,t)$ --- since the constraints are solved by the decomposition --- in agreement with our expectations from the classical theory. Thus, the Dirac quantization of the extended theory reproduces the known structures of finite dimensional quantum theories in the presence of global symmetries.

Even before we are able to compute the spectrum of $\hat H$, we can see how best matching has automatically produced much of the structure of the quantum theory. General solutions involve superpositions of eigenstates of the generalized momentum operators $\hat P_\alpha$ associated with the different manifest symmetries of the theory. The Ward-like identities arise naturally from the quantum constraints of the extended theory. The part of the structure of the Hilbert space and inner product is given by the spectrum of $\hat P_\alpha$. In the simple case where the configuration space represents particle motions in $\mathbbm R^d$ invariant under translations and rotations, the results of the last section suggest that the decomposition \eqref{eq:Psi decomp} represents the standard decomposition in terms of momentum and angular momentum eigenstates.

\subsubsection{Summary}\label{sec:Noether Summary}

We will now summarize the key results of this section regarding the properties of theories with manifest symmetries and fixed variations. These results confirm the statements made in Section~\ref{sec:gen argument} and will be important later.
\begin{itemize}
    \item In the classical theory, manifest symmetries with fixed variation imply the existence of conserved charges, $\pi^0_\alpha$, given by \eqref{eq:noether}. The value of the conserved charges is considered physical and is characteristic of the behaviour of the system in question. In practice, the conserved charges and are obtained from the initial conditions of the evolution.
    \item The classical observables correspond to functions on the \emph{full} unextended phase space $\Gamma(q^a,p_a)$.
    \item In the quantum theory, a general state is given by a \emph{superposition} of eigenstates of the symmetry generators according to the decomposition \eqref{eq:Psi decomp}.
    \item The quantum Dirac observables correspond to self-adjoint operators acting on $\mathcal H_\text{phys}$, which can be identified as the full Hilbert space of the \emph{unextended} theory. In particular, the operators associated with the auxiliary fields, $\hat \theta^\alpha$, are \emph{not} Dirac observables of the theory.
\end{itemize}

\subsection{Free Variations and Gauge Theories}\label{sec:free variations}

In this section, we will give a concrete definition of \emph{free} variation, as prescribed by best matching. Upon phase space reduction, this will lead to all the usual features of gauge theory including the appearance of Gauss law constraints on $\Gamma$ that effectively reduce the configuration space degrees of freedom by one, in agreement with the claim that a gauge symmetry indicates the presence of a non-physical degree of freedom in $\mA$. We will also see that the best matching procedure turns a global symmetry to a local one, as expected in gauge theory. However, from the best matching perspective, this is a secondary consequence of a deeper principle: the free variation, which is implied by the nature of the connection between the degrees of freedom and characteristic behaviour of the system, on the one hand, and the elements of $\mA$ and variational principle, on the other.

\subsubsection{Classical Theory: Gauss Constraints}\label{sec:Classical Gauge Theory}

We will now describe the conditions for free variation. The purpose of the extension procedure is to isolate the dependence of the theory on the coordinate in $\mA$ associated with the symmetry in question. For this purpose, we introduced the compensator fields $\theta^\alpha$. In a free variation, we require that the coordinate compensated by $\theta^\alpha$ is otiose and, therefore, cannot be specified by any physical initial conditions (it has no connection to the characteristic behaviour of the system the theory describes). In addition, we can guarantee that the otiose variable $\theta$ will have no effect on the predictions of our theory by requiring that the result of our variation be unchanged for \emph{any} choice of the location of the endpoints in the variation.\footnote{This additional requirement refines the motivations given in \cite{barbour_el_al:physical_dof} for what was referred to as a ``free \emph{endpoint} variation'', despite being a more restrictive requirement than standard free endpoint variation. This explains our terminology ``condition for free variation'' since our ``free variation'' is indeed distinct from standard ``free \emph{endpoint} variation''. For more details surrounding these issues, see \cite{barbour:scale_inv_particles,sg:dirac_algebra,and2010}.} This can be achieved by requiring, in addition to the standard Euler--Lagrange equations, the condition
\begin{equation}\label{eq:bm cond}
    \pi_\alpha \approx 0
\end{equation}
along \emph{all} points of the trajectory. The effect of this condition, can understood in terms of the Euler--Lagrange equations as requiring separately the conditions
\begin{align}
	\diby{L}{\theta} &= 0 & \diby{L}{\dot\theta} &= 0.
\end{align}
These conditions clearly achieve our goal of completely removing the dependence of the theory on $\theta$. In terms of the Hamiltonian, the condition for free variation \eqref{eq:bm cond} can be applied as a constraint, which we will call the \emph{best--matching constraint}. If the best-matching constraint is satisfied, the theory is said to be background independent with respect to this symmetry.

Using the best-matching constraint, we can rewrite $C_\alpha$ in terms of the equivalent constraint $C_\alpha = p_a \Tind \alpha a b (\theta) q^b$. The resulting system
\begin{equation}
    \{ H(q^a,p_a);\, \pi_\alpha;\, p_a \Tind \alpha a b (\theta) q^b \}
\end{equation}
is automatically first class in a manifest theory because: 
\begin{enumerate}
\item The manifest property of the symmetry gives
\begin{equation}
    \pb{\pi_\alpha}{H(q^a,p_a)} = 0;
\end{equation}
\item We have that $H(q^a,p_a)$ and $L_\alpha$ are first class because they can be obtained from a canonical transformation of a first class system, as seen in Section~\ref{sec:classical fixed};
\item Since $\pb{L_\alpha}{L_\beta} \approx 0$ (from point 2) we also have that:
\begin{equation}
    \pb{\pi_\alpha}{p_a \Tind \beta a b q^b} \approx \frac 1 2 f\indices{_{\alpha\beta}^{\gamma}} p_a \Tind \gamma a b q^b \approx 0
\end{equation}
where $f\indices{_{\alpha\beta}^{\gamma}}$ are the structure constants of $\mathfrak g$. 
\end{enumerate}
If the symmetry is \emph{not} manifest, then point 1 will not hold in general. However, if the Hamiltonian is proportional to a constraint, then there may be still be a \emph{hidden} symmetry in the theory and the procedure may still be well defined. This possibility will be explored in Section~\ref{sec:Gravity} in the context of gravity.

Since we have two first class constraints on the extended phase space, we expect these to constrain $\theta^\alpha$ and $\pi_\alpha$ as well as an additional $2\dim(\mathcal G)$ degrees freedom on $\Gamma_\text{e}$. To see that these degrees of freedom correspond precisely to the symmetry directions of $\mA$ (and their associated momenta), we can perform a phase space reduction from $\Gamma_\text{e} \to \Gamma$.

The simple gauge fixing $\theta^\alpha\approx 0$ trivially gauge fixes $\pi_\alpha \approx 0$. For the same reasons as before, the Dirac bracket reduces to the Poisson bracket after the phase space reduction. Applying the constraints strongly is straightforward. The gauge fixing $\theta^\alpha = 0$ sends $\Tind \alpha a b \to \tind \alpha a b$ and $\pi^\alpha = 0$ has no effect. The remaining system consists of the invariant Hamiltonian and the Gauss law constraints
\begin{equation}\label{eq:Gauss}
    P_\alpha = p_a \tind \alpha a b q^b \approx 0.
\end{equation}
Using the symplectic structure on $\Gamma(q^a,p_a)$, it is easy to see that these constraints generate the infinitesimal symmetry transformations of $\mathcal G$. The Dirac observables of this theory are, therefore, given not by arbitrary functions on $\Gamma$, as in the case of a free variation, but only functions that commute with the Gauss constraints. This difference between the definition of observables in the cases of theories with free or fixed manifest symmetries will be essential to the discussion of Section~\ref{sec:Formal Constructions}. The null vector fields generated by the Gauss constraints on phase space  correspond to the symmetry directions of $\mA$. The fact that functions which vary in these directions cannot be associated with observable quantities is thus consistent with the original requirement that these directions are not physical, justifying our procedure.

The Gauss constraints \eqref{eq:Gauss} can be compared with the identities \eqref{eq:noether}. The identities \eqref{eq:noether} express that there is a conserved charge in the theory that should be determined by the initial conditions. In contrast, the Gauss constraint fixes this charge to exactly zero. Because there is no physical input to fix the initial conditions of the coordinate associated with this symmetry, the theory must fix it internally. Gauss constraints thus owe their origin to surplus structure within our formalism, and not in any feature of the physical system being described. The difference is exactly in the number of independently specifiable initial data determined by the characteristic behaviour of the system.

We make one final observation: after inverting the Legendre transform, $ L \to  L(q^a, \dot q^a + N^\alpha\tind \alpha a b q^b)$, where $N^\alpha$ is a Lagrange multiplier for the Gauss constraint \eqref{eq:Gauss}. This is consistent with the usual picture of promoting a global symmetry to a local one by exchanging derivatives with covariant derivatives on a gauge bundle. Best matching reproduces this as a consequence of the free variation, which is taken to be a more fundamental principle.

\subsubsection{Quantum Theory: Dirac Constraints}

It will be useful to illustrate certain features of a quantum theory with free variation to compare it to the quantization of a fixed-variation theory. Unlike the fixed case, the quantization of the extended theory does not add any particular insight to the quantization of the gauge-fixed theory presented above. Thus, we will present only the quantization of the theory with $\theta^\alpha$ eliminated.

Dirac quantization requires that the Poisson structure be promoted to the commutator structure $\pb{q^a}{p_b} = \delta^a_b \to [\hat q^a, \hat p_b]\Psi = i\hbar \delta^a_b \Psi$ for operators acting on the elements, $\Psi$, of the Hilbert space, $\mathcal H$. The physical Hilbert space, $\mathcal H_\text{phys}$, is defined by states satisfying the quantum constraints
\begin{equation}\label{eq:quantum Gauss}
    \hat P_\alpha \Psi = 0.
\end{equation}
Time evolution is given by
\begin{equation}
    \hat H \Psi = i\hbar \diby \Psi t.
\end{equation}
We can now use the same conventions as before and take a configuration basis for $\mathcal H$ and the operator ordering \eqref{eq:P ordering} for $\hat P_\alpha$. The Gauss constraint \eqref{eq:quantum Gauss} selects the eigenfunction $\psi_0$ with eigenvalue $\pi^0_\alpha = 0$. The wavefunction is, thus, given by
\begin{equation}
    \Psi(q^a, t) = \psi_0(q^a,t).
\end{equation}

This illustrates the main difference at the quantum level between a fixed and free variation: in a fixed variation, $\Psi(q^a,t)$ can be in a superposition of eigenstates of $\hat P_\alpha$ while, in a free variation, $\Psi(q^a,t)$ is constrained to a single eigenstate of $\hat P_\alpha$. This additional restriction is reflected in the Dirac observables of the quantum theory. Whereas, in the case of fixed variations, they are simply arbitrary self-adjoint operators acting on $\mathcal H$, for free variations, they must also commute with $\hat P_\alpha$. This is equivalent to their definition as arbitrary self-adjoint operators acting on $\mathcal H_\text{phys}$. Crucially, the extra restriction derives from the fact that the coordinate associated to the symmetry in question is unphysical. We will return to  the origin and implications of these two differing notions of quantum observables in the specific case of reparametrization invariant theories in Section~\ref{sec:Dirty Quant}.

\subsubsection{Summary}\label{sec:gauge summary}

We will now highlight the key results of this section to compare with those of Section~\ref{sec:classical fixed}.

\begin{itemize}
    \item In the classical theory, manifest symmetries with fixed variation imply Gauss constraints \eqref{eq:Gauss}, which indicate that one of the coordinates of $\mA$ is unphysical. The Gauss constraints replace the Noether identities \eqref{eq:noether} because no physical information is available to fix the value of the conserved charge.
    \item The set of classical observables is resultantly smaller than that for a fixed variation theory by exactly the dimension of the relevant symmetry group. This indicates that the associated coordinates are merely facets of the formalism, not the physics, and justifies their classification as gauge degrees of freedom.
    \item In the quantum theory, a general state \emph{cannot} be in a superposition of eigenstates of $\hat P_\alpha$. It must be in the eigenstate with eigenvalue zero.
    \item The set of Dirac observables of the quantum theory are correspondingly smaller than the observables of a fixed variation theory in accordance with the classical observables. 
\end{itemize}

\subsection{General Prescription for Quantization }
\label{sec:Quant Prescrip}

The material presented provides a basis for the following prescription for the quantization of a finite dimensional theory with manifest symmetries.   

\begin{enumerate}
    \item Given $\mA$ and $S[\gamma]$, determine whether there exists a manifest symmetry associated with a particular degree of freedom.

    \item Given an identification of the characteristic behaviour of the system, determine the nature of the variational principle --- either fixed or free --- to be applied to the degree of freedom in question.

    \item Perform the Legendre transform of the system, extend the phase space by including the auxiliary pair $(\theta^\alpha,\pi_\alpha)$, and append the appropriate constraints.
     
    \item Determine a transformation of the variables and constraints that isolates the degree of freedom or symmetry in question.
    
    \item If the degree of freedom is associated with a fixed variation, proceed directly to the quantum theory via standard Dirac quantization.\footnote{Here we again note that, with our prescription, the Dirac quantization methodology can be substituted for a number of more rigorous modern approaches to the quantization of gauge theories provided our classification of degrees of freedom and constraints is preserved. See \S3.2.2 for references.} 
    
    \item If the degree of freedom is associated with a free variation, add the free variation condition $\pi_\alpha \approx 0$. Proceed to the quantum theory via Dirac quantization. (Note: the Dirac analysis will indicate whether there is a manifest or hidden symmetry, or whether the theory is inconsistent.)
\end{enumerate}

\subsection{A Note on Field Theories}

The considerations above are sufficient for the main case of interest in this paper, which is that of gravity. However, the class of finite dimensional models considered above is quite restrictive, so one might wonder about the general applicability of these concepts to more general fields theories like Maxwell or Yang--Mills. Detailed consideration of these cases in the context of our classification scheme would require a separate treatment and would, moreover, distract from our principal goal of better  understanding symmetry and evolution in quantum gravity. However, we believe that most of the key features described here will generalize to more sophisticated theories. Let us, therefore, take a brief moment to sketch out how.

In known field theories exhibiting (manifest) gauge symmetries --- such as Maxwell, Yang--Mills, and GR --- a well-defined classical evolution can only be achieved by finding initial data that obey the canonical constraints (such as the Gauss or Diffeomorphism constraints). This precisely corresponds to a decomposition of the initial data of the system into those that correspond to the propagating degrees of freedom (i.e., those that obey the initial value constraints) and the non-propagating modes (i.e., those that are eliminated by the constraints). This effectively implements a fixed variation for the propagating modes and a free variation for the non-propagating gauge modes, in accordance with our classification scheme. In the quantum theory, this translates into the requirement of constructing a physical Hilbert space on which the classical constraints are promoted to operator constraints on the wavefunction. Indeed, it is an often overlooked point that the spacelike boundaries of the path integral for gauge fields, (e.g., $A^\mu(x)$) should integrate over \emph{all} values of the pure-gauge modes, as in a free variation, while keeping the remaining components fixed.\footnote{This is, for example, why ghost fields have no external legs.}

Furthermore, the presence of timelike boundary conditions is known to generically break gauge invariance. In this case, one has the choice of adding additional boundary terms to restore gauge invariance, as in a free variation, or fixing boundary conditions inducing charges on the boundary, as in a fixed variation. These features seem to be in complete analogy to the considerations above. However, a more detailed analysis would be required to show this more carefully.

\section{Reparametrization Invariance in Finite Dimensions} \label{sec:Formal Constructions}

In this section, we consider the case of reparametrization invariant theories in which the temporal labeling applied to sequences of configurations is unphysical. In our terminology, reparametrization transformations should be understood as manifest symmetries of these theories. This much is uncontroversial. The crucial question is whether the relevant variation should be understood as fixed or free. This depends upon whether there exist degrees of freedom upon which the variational principle places no restriction: unphysical variables, with no connection to the characteristic behaviour of the system being described. First, in Section~\ref{sec:Rep Class}, we will give a general demonstration that, in finite dimensions, the variational principle for reparametrization invariant theories is of the fixed type. This  implies that reparametrization should be treated as a conservation symmetry, rather than a gauge-type symmetry. We will then see, in Section~\ref{sec:Dirty Quant}, that this reclassification leads to a relational quantum formalism importantly different to that researched within conventional gauge theories treatments. The lessons learned in this analysis will prove crucial when we pass the the full relativistic formalism in Section~\ref{sec:Gravity}.         

\subsection{Reparametrization as a \textit{Manifest} Symmetry of \textit{Fixed} Variations}
\label{sec:Rep Class}

Let us return to our original characterization of manifest symmetries in finite dimensions. There we considered a configuration space, $\mA$, with coordinates on which a Lie group acts such that the induced coordinate transformation was a global invariance of the action. Although reparametrization \textit{is} a Lie group, unlike the groups we considered earlier it has no associated action on $\mA$. Rather, since it induces a mapping between different parametrizations of \textit{the same} configuration space curve, we should think about its action as being that of the one dimensional real diffeomorphism group, $\mathcal{D}iff(\mathbb{R})$, upon the space of parameterized configuration space curves, $\mathcal{C}_{p}\in \{\gamma,t\}$ made up of a pairing of a curve, $\gamma$, and a monotonic parameter, $t$. For the finite dimensional case, the action of the reparametrization group, $\mathcal{R}ep$, takes the simple form:
\begin{equation}
\{ \gamma,t \} \rightarrow \{ \gamma,f(t) \}
\end{equation}
for $\frac{df(t)}{dt}>0$. Significantly, if we project the group action back onto $\mA$ then we just get the identity. As one would expect, reparametrization just gives us the same sequence of configurations. This will not be the case if we consider the velocity-configuration space $T\mA$, since the velocities depend upon the time parametrization. It is therefore useful to think of the consequence of $\mathcal{R}ep$ on $T\mA$:
\begin{equation}
\Big\{q^{a},\frac{dq^{a}}{dt} \Big\}\rightarrow \Big\{q^{a},\frac{df}{dt}\frac{dq^{a}}{dt}\Big\}
\end{equation} 
If we consider two parametrization of the same sequences of configurations we will thus find two distinct curves in velocity-configuration space. The crucial hallmark of reparametrization invariant theories is that the relevant Lagrangian transforms such that, $\dot{q}^{a}\rightarrow \frac{df}{dt} \dot{q}$ implies $L[q^{a },\dot{q}^{a}]\rightarrow \frac{df}{dt} L[q^{a },\dot{q}^{a}]$. This means that the action of the theory will be invariant under rescalings of the parameter $t$. Thus, if we consider two parametrizations of the same sequences of configurations, we will find two distinct curves on $T\mA$, each corresponding to the same value of the action. This fits our definition of a \textit{manifest} symmetry above. 

Explicitly, following Dirac \cite{dirac:lectures}, we have that since the Lagrangian is  homogeneous of degree one in the velocities $\dot{q}^{a}$ (i.e., we have that for some $k$ the transformation $\dot{q}^{a}\rightarrow k \dot{q}$ implies $L[q^{a },\dot{q}^{a}]\rightarrow k L[q^{a },\dot{q}^{a}]$) by Euler's homogeneous function theorem:
\begin{equation}
\dot{q}^{a }\frac{\partial L}{\partial \dot{q}^{a}}=L 
\end{equation}
which implies 
\begin{equation}
L-\dot{q}^{a }\frac{\partial L}{\partial \dot{q}^{a}}=p_{a}\dot{q}^{a }- L=H=0.
\end{equation}
Thus, the Hamiltonian must be proportional to a constraint.

There are two immediate and important atypical aspects of exactly how reparametrization symmetry works within the Hamiltonian formalism. First, the constraint obtained as a result of the non-invertibility of the Legendre transform also acts as the Hamiltonian of the system. Consequently, its flow generates the dynamics of the system, which must be parametrized by a monotonic parameter. Thus, the Lagrange multiplier of this constraint is restricted to be positive definite, unlike standard Lagrange multipliers.

Second, given some initial point in phase space, the dynamics is given by a \emph{unique} phase space curve originating from this point. Thus, there is \emph{no} under-determination problem on phase space. This is because, by definition, the momenta are such that they are insensitive to the action of reparametrizations on velocity-configuration space:
\begin{equation}
p_{a}=\frac{\partial L}{\partial \dot{q}^{a}}\rightarrow \frac{\partial L}{\partial \dot{q}^{a}}\frac{df}{dt}\frac{dt}{df}=p_{a}.
\end{equation}
This implies that any two points in $T\mA$ which differ by the action of a reparametrization will be mapped by the Legendre transformation to the same point in phase space. Thus, phase space curves will be isomorphic to the original curves in $\mA$ and the variational principle in phase space will be completely determined by fixing the initial data.
           
For our symmetry classification scheme, the most significant detail is now whether the variation should be understood as free or fixed. Clearly, the only option is \emph{fixed} since the variation is with respect to a variable which is not a phase space degree of freedom! On phase space, the variational principle is \textit{already} well defined and {\it all directions in $\mA$ are parameterized by degrees of freedom with a direct correspondence to the characteristic behaviour of the system in question.}

\subsection{Relational Quantization Procedure}
\label{sec:Dirty Quant}

We have shown that reparametrization invariance is a manifest symmetry associated with a fixed variation and, thus, it should be treated in the manner of a conservation symmetry, where the conserved charge is then interpreted as the total energy of the system. Classically, although this result is conceptually  interesting, it has no empirical implications. Quantum mechanically however, the result is non-trivial since, as we shall see, it leads us to a substantially different quantum theory. 

Following the prescription detailed in Section~\ref{sec:Quant Prescrip} (and noting that we have already completed steps 1 and 2), we make the phase space extension $\Gamma(q,p)\rightarrow \Gamma_{e}(q,\tau;p,\epsilon)$, where we have labeled the single auxiliary configuration variable $\tau$ and its momenta $\epsilon$. Following the prescription given in \cite{gryb:2011}, we identify the constraint
\begin{equation}\label{eq:rel quant}
    H_{e}=H+\epsilon
\end{equation}
as the constraint restricting the auxiliary fields $(\tau,\epsilon)$. This choice is motivated by the observation that the generator of the symmetry transformation is the Hamiltonian itself. Thus, the analogue of $L_\alpha$ for reparametrization symmetries is $H$. With this observation, it is clear that the generalization of $C_\alpha = L_\alpha - \pi_\alpha$ is $H_e$ given in \eqref{eq:rel quant}. Furthermore, as argued in \cite{gryb:2011}, this corresponds to an \textit{equitable} --- in the sense of depending upon all the degrees of freedom --- choice of internal clock after a deparametrization with respect to $\tau$. The crucial property of $H_e$ for what follows is that $H(q,p)$ is independent of both $\tau$ or $\epsilon$. Thus, $H_e$ can be thought of as generating \emph{time-independent} evolution in terms of an arbitrary label $\tau$.

Now, since we have identified reparametrization as being of the Case~2 type --- i.e., a manifest fixed symmetry --- we do \textit{not} impose the further constraint $\epsilon=0$ (to do so would lead back to the conventional `timeless'  Wheeler--DeWitt-type analysis). Rather, we  proceed to the quantum theory via the standard Dirac method of promoting the constraint to an operator constraint on the wavefunction --- i.e., we have 
\begin{equation}\label{eq:RQ phys H}
   \hat{H}_e \ket\Psi_{\text{phys}} = 0,
\end{equation}
with
\begin{equation}
   \hat{H}_e = \hat{H} + \hat{\epsilon},
\end{equation}
as the definition of the states, $\ket\Psi_{phys}$, that are elements of the physical Hilbert space $\mathcal{H}_{phys}$.\footnote{Here we note, once more,  that there are well known formal issues with Dirac quantization that render such a Hilbert space --- strictly speaking --- not well defined. In a fully rigorous application of relational quantisation to a  reparametrization invariant system more powerful techniques, such as group averaging \cite{Giulini:1999a,Giulini:1999b}, would need to be used. Such modifications would not imply any difference in the basic structure of our arguments.} If we pick a simple basis for the operators $\hat\tau$ and $\hat{\epsilon}$:
\begin{align}
    \hat\tau &= \tau & \hat {\epsilon} &= -i\frac{\partial}{\partial\tau},
\end{align}
then we can recover the normal Heisenberg evolution equation for quantum theory. The above basis is equivalent to deparametrizing the extended classical theory with respect to $\tau$ and then quantizing. Observables are now taken to be represented by self-adjoint operators acting on the physical Hilbert space. The role of time is now played by an operator corresponding to the arbitrarily determined evolution parameter:
\begin{equation}  
[\hat{H},\hat{o}]= -i\frac{\partial\hat{o}}{\partial\tau}.
\end{equation}
This formalism leads us to ask what interpretation we should ascribe to the evolution parameter. Unlike the time parameter of conventional quantum theory, in our relational formalism it is an operator. This might be thought to imply that, in this approach, $\tau$ is itself a physical observable associated with a clock. However, by construction we have that $\tau$ is not an operator on the physical Hilbert space. Rather  it is an operator on the space of auxiliary states, $\ket\Psi_{\text{aux}}$, which are defined prior to the imposition of the constraint condition \eqref{eq:RQ phys H}. Time is thus explicitly not an observable of the theory. Rather, according to this formalism, what \textit{is} observable are the successive, distinct values of each genuine physical degree of freedom. Nevertheless, the role of time within a relationally quantized theory is clearly a novel one: it is no longer a classical background, but an \textit{unobservable} quantum operator. The possibly radical  interpretational consequences of this change are are an interesting avenue of ongoing investigation. 

In the Schr\"odinger picture, we can compare the results of Relational Quantization with the results of a straightforward Dirac quantization of reparametrization invariant theories. In Relational Quantization, the physical Hilbert space is defined by \eqref{eq:RQ phys H}. We can solve this constraint explicitly if the spectrum of $\hat H$ is known using the ansatz
\begin{equation}
    \ket\Psi_\text{phys} = \sum_n c_n e^{-iE_n \tau} \ket{\psi_n},
\end{equation}
where $\ket{\psi_n}$ are the eigenvalues of $\hat H$ with eigenvalue $E_n$. Thus, as is consistent with the treatment of conservation symmetries (as summarized in Section~\ref{sec:Noether Summary}), the general solution corresponds to a superposition of eigenstates of the symmetry generator. In contrast, the conventional treatment leads to a physical Hilbert space, $\mathcal H_\text{Dirac}$, defined by the operator constraint equation
\begin{equation}
    \hat H \ket\Psi_\text{Dirac} = 0.
\end{equation}
This is solved explicitly by
\begin{equation}
    \ket\Psi_\text{Dirac} = \ket{\psi_0}.
\end{equation}
Thus, the Universe is trapped in an energy eigenstate and the evolution is frozen. This treatment is consistent with associating reparametrization invariance with some conventional gauge parameter (as summarized in Section~\ref{sec:gauge summary}), typically understood to be an internal clock.

We have therefore carved out a middle ground, hitherto unoccupied, whereby our quantum formalism is neither eternally frozen, nor admits an absolute temporal background. A forthcoming paper will be devoted to the further investigation of the formal properties of the observables defined by our procedure. In particular, we will demonstrate the close formal correspondence between these observables and the observables first proposed by Kucha\u r \cite{kuchar:1981,Kuchar:1993}  (see also \cite{torre:1993,hajicek:2000,Barbour:2008,Anderson:2012}. We will now turn our attention to the gravitational field and show how our relational quantization can be applied in the context of General Relativity.
 
\section{Gravity} \label{sec:Gravity}

York's ontology, as discussed in Section~\ref{sec:York Ontology}, combined with the general method for treating symmetries presented in Section~\ref{sec:Quant Prescrip}, leads unambiguously to a concrete proposal for the evolution equation of the wavefunction of the Universe consistent with the Relational Quantization methodology described in Section~\ref{sec:Formal Constructions}. In this section, we show how this is achieved. We begin the section by reviewing standard ADM theory \cite{ADM:1960}.

\subsection{Symmetry in ADM}\label{sec:ADM}

The ADM theory contains \emph{simultaneous} free variations with respect to infinitesimal spatial diffeomorphisms and local reparametrizations. To see this, consider the ADM action in $d+1$ dimensions (for $d>2$):
\begin{equation}
        S_\text{ADM} = \frac 1 {2\kappa} \int_{\mathcal{M}} d^d x\, dt\sqrt g \lf[ \frac 1 {N} K_{ab} G^{abcd} K_{cd} + N \lf( R(g) - \frac{d(d-1)k}{\ell^2} \rt) \rt],
\end{equation}
where, $g_{ab}$ is the spatial metric,
\begin{equation}
    K_{ab} = \dot g_{ab} + \lie_{N^a} g_{ab}
\end{equation}
is the extrinsic curvature, $N$ is the \emph{lapse}, $N^a$ is the \emph{shift}, $R(g)$ is the spatial curvature, and
\begin{equation}
    G^{abcd} = g^{ac}g^{bd} + \lambda g^{ab}g^{cd}
\end{equation}
is the DeWitt supermetric. For the couplings of the theory, we have that $k = \frac 1 {8\pi G}$ is the gravitational constant, $\lambda$ takes its ADM value of $-1$, $\ell$ is the Hubble radius, and $k = 0, \pm 1$ so that
\begin{equation}
    \Lambda = \frac {d(d-1)k} {2\ell^2}
\end{equation}
is the cosmological constant. We restrict to globally hyperbolic Lorentzian manifolds of topology $\mathcal M = \Sigma \times \mathbbm R$, where $\Sigma$ is a $d$ dimensional Euclidean compact manifold without boundary. We have kept the dimensionality of space arbitrary in order to keep our expressions as general as possible.

We now note the following:
\begin{itemize}
    \item The action of the infinitesimal spatial diffeomorphisms on $g_{ab}$ is given by
    \begin{equation}
	g_{ab} \to g_{ab} + \epsilon \lie_{\xi} g_{ab} = g_{ab} + \epsilon \lf( \nabla_a \xi^c g_{bc} + \nabla_b \xi^c g_{ac} \rt) ,
    \end{equation}
    where $\epsilon << 1$.\footnote{A full treatment of spatial diffeomorphisms would require large diffeomorphisms, adding complications that don't affect our main argument. For simplicity, we will restrict our discussion to infinitesimal diffeomorphisms.} Since $S_\text{ADM}$ is invariant under time-independent (i.e., foliation preserving) diffeomorphisms, the extrinsic curvature $K_{ab}$ --- which contains all dependence of the time derivative of $g_{ab}$ --- has precisely the form required for a covariant derivative of $g_{ab}$ on a $\text{Riem}/\text{Diff}$-bundle, where Riem is the space of Euclidean metrics on $\Sigma$, Diff is the diffeomorphism group, and $\dot \xi$ is the connection (see \cite{Gomes:gauge_theory_riem} for a precise statement of this). If we identify the shift with
    \begin{equation}\label{eq:shift}
	N^a = \dot \xi^a
    \end{equation}
    then we can interpret $\xi$ as a compensator field for infinitesimal spatial diffeomorphisms.
    
    \item Local reparametrization invariance can be explicitly parametrized by an auxiliary field $\tau$ if we make the replacement
    \begin{equation}\label{eq:lapse}
	N = \dot \tau.
    \end{equation}
    
\end{itemize}

    By making the identifications \eqref{eq:shift} and \eqref{eq:lapse}, we have parametrized the spatial diffeomorphisms and local reparametrizations in a form suitable for treatment with Best Matching. We define the momenta
    \begin{align}
	\pi^{ab} &= \ddiby{S_\text{ADM}}{\dot g_{ab}(x)} & \chi_a(x) &= \ddiby{S_\text{ADM}}{\dot \xi^a(x)} & p_\tau(x) &= \ddiby{S_\text{ADM}}{\dot \tau(x)},
    \end{align}
    which implies the symplectic structure
    \begin{align}
	    \pb {g_{ab}(x)}{\pi^{cd}(y)} &= \delta^{cd}_{ab}(x,y) & \pb{\xi^a(x)}{\pi_b(y)} &= \delta^a_b(x,y) & \pb{\tau(x)}{p_\tau(y)} = \delta(x,y)
    \end{align}
    on the extended phase space $\Gamma_\text{e}(g_{ab}, \xi^a, \tau; \pi^{ab}, \chi_a, p_\tau)$. A short calculation, which uses integration by parts and the fact that $\partial \Sigma = 0$ for the second constraint, leads to the primary constraints $\ham$ and $V_a$ given by
    \begin{align}
	\ham &= 2p_\tau + \kappa \frac{\pi^{ab} G_{abcd} \pi^{cd}}{\sqrt g} - \frac 1 \kappa \lf(R- \frac{d(d-1)k}{\ell^2}\rt)\sqrt g \approx 0 \\
	V_a &= \chi_a + 2 g_{ab} \nabla_c \pi^{bc} \approx 0.
    \end{align}
    The ADM theory can be obtained by applying free variations to $\xi^a$ and $\tau$ such that
    \begin{align}
	\chi_a &\approx 0 & p_\tau &\approx 0.
    \end{align}
    These constraints can be trivially gauge fixed and thrown away using the gauge fixing conditions $\xi^a = 0$ and $\tau = t$. The remaining system of constraints,
    \begin{subequations}\label{eq:ADM constraints}
    \begin{align}
	\ham &= \kappa \frac{\pi^{ab} G_{abcd} \pi^{cd}}{\sqrt g} - \frac 1 \kappa \lf(R- \frac{d(d-1)k}{\ell^2}\rt)\sqrt g \approx 0 \label{eq:ADM Ham}\\
	V_a &= 2 g_{ab} \nabla_c \pi^{bc} \approx 0,
    \end{align}
    \end{subequations}
    is that of ADM on the ADM phase space $\Gamma_\text{ADM}(g_{ab};\pi^{ab})$. This system is known to be first class and obey the Dirac algebra \cite{adm:adm_review}
    \begin{align}
	\pb{\ham (N_1)}{\ham(N_2)} &= V(\zeta^a(N_1,N_2,g^{ab})), \label{eq:H-H pb}\\
	\pb{V(N^a)}{\ham(N)} &= \ham(\lie_{N^a} N), \label{eq:V-H pb}\\
	\pb{V(N^a_1)}{V(N^a_2)} &= V(\lie_{N^a_1} N^a_2),\label{eq:V-V pb}
    \end{align}
    where $\ham(N) := \int_\Sigma d^dx N(x) \ham (x)$, $V(N^a):= \int_\Sigma d^dx N^a(x) V_a(x)$, and
    $$\zeta^a(f_1,f_2,g^{ab}) = g^{ab} \lf( f_1 f_{2;b} - f_2 f_{1;b} \rt)$$
    for some smearing functions $f_1$ and $f_2$.\footnote{That the constraints \eqref{eq:H-H pb} and \eqref{eq:V-V pb} close only when the variations are free is a result of only considering infinitesimal, and not large, diffeomorphisms.} This algebra is known to generate hypersurface deformations \cite{Teitelboim:DT_algebra,Thiemann:book} that, when applied to a solution of the Einstein equations, corresponds to an infinitesimal spacetime diffeomorphism of $\mathcal M$ (that preserves the space-like embedding of the hypersurfaces). The fact that this symmetry is a manifest symmetry of the ADM action means that, because the variations of $\tau$ and $\xi^a$ are free, the ADM theory na\"ively falls into the first category of Figure~\ref{fig:table of dof}: that of a gauge symmetry. However, there are at least two problems with this classification:
    \begin{enumerate}
	\item The diffeomorphisms appear to act like the generators of a standard gauge symmetry, but they are not. Known gauge theories can be obtained either from making a global symmetry of the action local or by unveiling a hidden symmetry using a symmetry trading procedure. This implies that there exists some sort of consistent Hamiltonian flow on the unreduced phase space corresponding to the theory with the global symmetry (possibly after symmetry trading). This is not possible with the diffeomorphism constraints because the Poisson bracket \eqref{eq:H-H pb} implies that such a theory would be inconsistent: the Hamiltonian constraints $\ham$ \emph{require} the presence of the diffeomorphism constraints $V_a$ in order for the constraint algebra to close. This obscures the identification of the physical degrees of freedom.
	
	\item The free variation of $p_\tau$ does not respect York's requirement that the mean of the York time be fixed in the variation. We will see how this arises in detail in Section~\ref{sec:SD} but, for now, we note that there is no obvious way to express the Hamiltonian constraint in a form
	\begin{equation}
	    \ham = p_{\tau'} + H,
	\end{equation}
	where $H$ is independent of some arbitrary time label $\tau'$ (which could be different from $\tau$). Only in this form would the theory be consistent with the Relational Quantization methodology laid out in Section~\ref{sec:Dirty Quant}. The most na\"ive attempt to achieve this would be to require $\tau$ to have a fixed variation by lifting the condition $p_\tau \approx 0$. However, this fails because the constraint algebra no longer closes since the Poisson bracket \eqref{eq:V-H pb} becomes
	\begin{equation}
	    \pb{V(N^a)}{\ham(N)} = \ham(\lie_{N^a} N) - 2 \int_\Sigma d^d x\, N^a N_{;a} p_\tau\not\approx 0, \\
	\end{equation}
    \end{enumerate}

The failure of this procedure to generate a clean decomposition of the physical degrees of freedom is naturally seen to be a result of trying to apply canonical techniques, which fundamentally break spacetime diffeomorphism invariance, to a spacetime diffeomorphism invariant theory. These difficulties are related to the many assets of the Problem of Time, which are discussed extensively in \cite{Isham:pot_review,and2010}. Standard quantization techniques require, in one form or another, an identification of the physical degrees of freedom of the theory. We will now show that there exists a decomposition of the ADM phase space that accomplishes this explicitly.

\subsection{York's Ontology}

\subsubsection{The Characteristic Behaviour of Gravity}

York's proposal requires a fixed variation of the conformal geometry and the mean value of the York time, $P = \frac 2 {dV_g} \int_{\Sigma} d^dx\, g_{ab} \pi^{ab}$, which is canonically conjugate to the volume $V_g$\footnote{Note that, in the quantum theory, taking $P$, which takes values in $\mathbbm R$, as the independently specifiable degree of freedom instead of $V_g$, which takes values in $\mathbbm R^+$, avoids the issue of having to deal with an operator valued in $\mathbbm R^+$.}
\begin{equation}
    V_g =\int_\Sigma d^dx\sqrt g.
\end{equation}
The subscript $g$ indicates that the volume is computed using the metric $g_{ab}$ (the utility of this notation will become clear shortly). For $P$ to be fixed on the boundary of the variation, we must have that its conjugate variable, the volume, be specifiable there. This means that only the conformal modes of the metric that do not contribute to the volume should be varied freely. Considering this together with the requirement that the coordinate dependent information should be removed from the metric, we conclude that we are searching for a gauge theory of spatial diffeomorphisms and \emph{volume preserving} conformal transformations where the York time is fixed in the variation.

The natural starting point for our analysis is the ADM action, which has the rather undesirable feature --- as far as York's proposal is concerned --- of neither having manifest invariance under volume preserving conformal transformations, nor being a conventional gauge theory (as we discovered in the last section) with respect to diffeomorphisms, nor varying the York time in fixed manner (as we shall see). Remarkably, the first two difficulties can be resolved by noting that the volume preserving conformal transformations are a hidden symmetry of the ADM action. The last difficulty can be dealt with using Relational Quantization. 

The task we set presently is to demonstrate that volume preserving conformal transformations are hiding in the ADM theory. The first step is to parametrize the volume preserving conformal constraints using the conjugate pair $(\phi(x),\pi_\phi(x))$ and the volume \emph{changing} conformal constraints (this is to ensure the fixed variation of $P$) using the pair $(\phi_0,\pi_0)$. These will eventually become compensator fields for the relevant symmetries. To do this, we first extend the phase space such that $\Gamma(g_{ab}; \pi^{ab}) \to \Gamma_\text{e}(g_{ab}, \phi; \pi^{ab}, \pi_\phi)$. A noteworthy point is that, while $\phi(x)$ is a local field and $\pi_\phi(x)$ is a density of weight one on $\Sigma$, $\phi_0$ and $\pi_0$ are simply (time-dependent) spatial constants. Thus, the symplectic structure on $\Gamma_\text{e}$ is appended by
\begin{align}
    \pb{\phi(x)}{\pi_\phi(y)} &= \delta(x,y) & \pb{\phi_0}{\pi_0} = 1.
\end{align}
To ensure that the new variables behave as auxiliary gauge degrees of freedom, we vary them freely using the first class constraints
\begin{align}
    D &\equiv \pi_\phi \approx 0 & D_0&\equiv \pi_0\approx 0.
\end{align}

Now we can perform a canonical transformation that explicitly parametrizes the symmetries discussed above. Denoting by over-bars those variables that are obtained after the canonical transformation, we find that the generating functional that accomplishes this task is
\begin{multline}
    F\lf[g_{ab}, \phi, \phi_0; \bar \pi^{ab}, \bar \pi_\phi,\bar \pi_0 \rt] = \int_\Sigma d^d x \lf[  g_{ab} \exp \lf\{ -\frac 4 {d-2} \lf(  \phi - \frac{(d-2)}{2d}\lf( \log \mean{e^{\frac {2d}{d - 2} \phi}}_{g} + \phi_0\rt) \rt) \rt\} \bar \pi^{ab}\rt. \\ \lf. + \phi\bar\pi_\phi  + \phi_0 \bar \pi_0  \rt].
\end{multline}
For ease of notation, we have introduced the \emph{mean} operator $\mean{\cdot}$, whose action on scalar and densities of weight one is
\begin{align}
    \mean{\text{scalar}}_g &= \frac 1 {V_g} \int_\Sigma d^dx (\text{scalar}) \sqrt g & \mean{\text{density}}_g &= \frac 1 {V_g} \int_\Sigma d^dx (\text{density}).
\end{align}
From now on, we will also use the variable $\lambda$ to abbreviate the rather lengthy expression in the exponential:
\begin{equation}
    \lambda =  \phi - \frac {(d-2)} {2d} \lf( \log \mean{e^{\frac {2d}{d- 2} \phi}}_{g} + \phi_0 \rt).
\end{equation}

Using $F$, we can find $g_{ab}$, $\phi$, and $\phi_0$ in terms of barred quantities by inverting
\begin{align}
    \bar g_{ab} &= \ddiby{F}{\bar\pi^{ab}} & \bar \phi &= \ddiby{F}{\bar\pi_\phi}& \bar \phi_0 &= \ddiby{F}{\bar\pi_0}
\end{align}
then use these expressions to determine $\pi_{ab}$, $\pi_\phi$, and $\pi_0$ using
\begin{align}
    \pi_{ab} &= \ddiby{F}{g_{ab}} & \pi_\phi &= \ddiby{F}{\phi}& \pi_0 &= \ddiby{F}{\phi_0}.
\end{align}
These manipulations yield
\begin{align}
    g_{ab} &= e^{\frac 4 {d-2}\bar\lambda}\bar g_{ab} & \pi^{ab} &= e^{-\frac 4 {d-2} \bar\lambda} \lf[ \bar\pi^{ab} - \lf( 1 - e^{\frac {2d}{d - 2} \bar\lambda} \rt) \frac{\mean{\bar\pi}_{\bar g}} d \bar g^{ab} \sqrt {\bar g} \rt]
\end{align}
for the ADM variables (where we have defined $\pi \equiv g_{ab} \pi^{ab}$). It is straightforward to verify that the volume and the mean of the York time transform as
\begin{align}
    V_g &= e^{ \phi_0} V_{\bar g} & P &= e^{-\phi_0} \bar P,
\end{align}
by construction, so that $\phi_0$ is the part of $\lambda$ that changes only the volume. The compensator fields transform like
\begin{align}
        \phi&= \bar \phi & \phi_0 &= \bar \phi_0 \\
        \pi_\phi &= \bar\pi_\phi - \frac 4 {d-2} \lf( \bar\pi - \mean{\bar\pi}_{\bar g} \rt)  & \pi_0 &= \lf( \frac{\bar\pi_0}{V_{\bar g}} - \frac 2 d \mean{\bar\pi}_{\bar g} \rt) V_{\bar g}.
\end{align}

It will be convenient to decompose $\pi^{ab}$ into a traceless part, $\sigma^{ab}$, and a trace part, $\pi$, such that
\begin{align}
    \sigma^{ab} &= \pi^{ab} - \frac 1 d \pi g^{ab} & \pi &= g_{ab} \pi^{ab}.
\end{align}
In terms of these variables, the ADM constraints \eqref{eq:ADM constraints} become
\begin{align}
    \ham &= \frac{\kappa}{\sqrt g} \lf( \sigma^{ab} \sigma_{ab} - \frac {\pi^2}{d(d-1)} \rt) - \frac{\sqrt g}{\kappa} \lf( R - \frac {d(d-1)k}{\ell^2} \rt) \\
    V_a&= 2 \lf( g_{ab} \nabla_c \sigma^{bc} + \frac 1 d \nabla_a \pi \rt).\label{eq:diff sigma}
\end{align}
The transformation properties of $\pi^{ab}$ can be expressed in an enlightening form in terms of these variables:
\begin{align}
    \sigma^{ab} &= e^{-\frac 4 {d-2}\bar \lambda} \bar \sigma^{ab} & \pi & = \bar \pi - \lf(1- e^{\frac {2d} {d-2}\bar\lambda}\rt) \mean{\bar\pi}_{\bar g} \sqrt {\bar g} \label{eq:T pi}
\end{align}
The traceless part transforms homogeneously while the trace part has an inhomogeneous piece. This inhomogeneous piece is crucial for the uniqueness properties that will follow, a fact first appreciated in \cite{barbour_el_al:physical_dof}.

\subsubsection{The Constraints}

The first class constraints
\begin{align}
    D = \pi_\phi &\approx 0 & D_0 = \pi_0 &\approx 0
\end{align}
are easily seen to transform to (after a simple renormalization of $D_0$ by $V_{\bar g}$)
\begin{align}
    D = \bar \pi_\phi - \frac 4 {d-2} \lf( \bar \pi - \mean {\bar \pi}_{\bar g}\rt) &\approx 0 & D_0 = \frac{\bar \pi_0}{V_{\bar g}} - \frac 2 d \mean{\bar \pi}_{\bar g} &\approx 0.
\end{align}
As hoped, $D$ expresses that $\bar \pi_\phi$ is equal to the generator of volume preserving conformal transformations while $D_0$ expresses that $\bar \pi_0$ is equal to the \emph{dimensionless} York time.

We will now show that the volume preserving conformal transformations are a hidden symmetry of the ADM action. Since the compensator fields $(\bar \phi,\bar \phi_0)$ enter only through $\bar\lambda$, it is convenient to write the canonical transformation as a map $T_{\bar \lambda}$ acting on $\Gamma_\text{e}$. Using this notation, the ADM constraints transform formally to
\begin{align}
    T_{\bar \lambda} \ham &\approx 0 & T_{\bar \lambda} V_a &\approx 0.
\end{align}
It is appropriate now to apply the free variation conditions
\begin{equation}\label{eq:phi gf}
    \bar \pi_\phi = 0
\end{equation}
required by the York ontology. These conditions gauge fix all but one of the scalar constraints $T_{\bar \lambda} \ham \approx 0$. That this is a valid gauge fixing condition will follow if these constraints can be solved for the variable, $\bar\phi$, conjugate to $\bar\pi_\phi$. Generally, for gauge fixings of this type, where the gauge fixing is itself a phase space variable, the Dirac bracket reduces to the Poisson bracket on the reduced phase space with $(\bar\phi,\bar\pi_\phi)$ eliminated because each additional term in the Dirac bracket will contain a Poisson bracket of a function on the reduced phase space with $\bar\pi_\phi$, which is zero by definition (see \cite{Dirac:CMC_fixing} for a more careful illustration of this).

To perform the gauge fixing, we apply the constraints strongly and solve for the remaining first class constraint. This can be done by noting that \eqref{eq:phi gf} implies $\bar \pi \approx \mean{\bar\pi}_{\bar g} \sqrt {\bar g}$ so, by \eqref{eq:T pi}
\begin{equation}
    \pi \approx \mean{\bar \pi}_{\bar g} e^{\frac {2d}{d- 2}\bar\lambda} \sqrt{\bar g}.
\end{equation}
We can define an equivalent constraint surface by combining this with the $\pi^2$ term of the previous $T_{\bar\lambda}\ham$ and obtain
\begin{multline}\label{eq:LY eqn}
    T_{\bar\lambda}\ham \to T_{\bar \lambda}\ham = \frac{\kappa\bar\sigma^{ab} \bar\sigma_{ab}}{e^{\frac {2d}{d- 2}\bar\lambda}\sqrt{\bar g}}  + \lf( \frac{d(d-1)k}{\kappa \ell^2} - \frac{\kappa \mean{\bar \pi}_{\bar g}^2}{d(d-1)}  \rt)e^{\frac {2d}{d- 2}\bar\lambda}\sqrt{\bar g} \\ - \lf( R(\bar g) - \frac{4(d-1)}{d-2} \bar g^{ab} \lf( \bar\lambda_{;ab} + \bar\lambda_{;a} \bar\lambda_{;b} \rt) \rt) \frac{e^{2\bar\lambda} \sqrt{\bar g}}\kappa \approx 0.
\end{multline}
Equation~\eqref{eq:LY eqn} is the Lichnerowicz--York equation for $\bar\lambda$. Solutions to this equation are know to exist,\footnote{Uniqueness depends on the value of the cosmological constant and the matter content of the theory. For details, see \cite{Gomes:matter_coupling_proc,Gomes:h_gl_uniqueness}.} \cite{Niall_73} justifying our claim that this is a valid gauge fixing and our use of the gauge-fixing conditions as strong equations. We can re-write the remaining constraint, not gauge fixed by \eqref{eq:phi gf}, using the definition of $\bar\lambda$. This yields
\begin{equation}
    \boxed{e^{\bar\phi_0} \approx \mean{e^{\frac {2d}{d- 2}\lambda_0}}_{\bar g}}.
\end{equation}
where $\lambda_0$ is given by
\begin{equation}
   \boxed{ \lf. T_{\bar\lambda} \ham \rt|_{\bar\lambda = \lambda_0} = 0}.
\end{equation}
The conformal constraints immediately become
\begin{equation}\label{eq:vpc const}
    \boxed{D = \bar \pi - \mean{\bar \pi}_{\bar g} \sqrt {\bar g} \approx 0}.
\end{equation}
Because these constraints generate volume preserving conformal transformations, we reach the important conclusion that the free variation of the conformal factor of metric, as required by the York ontology, makes the volume preserving conformal transformations a manifest symmetry of the theory.

After gauge fixing, the diffeomorphism constraints can be handled straightforwardly. We make the key observation that $\nabla_a \mean{\pi} = 0$, first noticed by York, to re-write $V_a$ in \eqref{eq:diff sigma} as:
\begin{equation}
    V_a = 2 \lf( g_{ab} \nabla_c \sigma^{bc} + \frac 1 d \nabla_a \lf( \pi - \mean{\pi}_g \sqrt g \rt) \rt) \approx 0.
\end{equation}
From \eqref{eq:T pi}, it follows
\begin{equation}
    \pi - \mean{\pi}_g \sqrt g = \bar \pi - \mean{\bar \pi}_{\bar g} \sqrt {\bar g}.
\end{equation}
A short calculation shows that the Christofel symbols $\Gamma^c_{ab}$ transform as
\begin{equation}
    \Gamma^c_{ab} = \bar \Gamma^c_{ab} + \frac 2 {d-2} \lf( \delta^c_{(a} \partial^{\phantom{c}}_{b)} \bar\lambda -  \bar g^{dc} \bar g_{ab} \partial_d \bar \lambda  \rt).
\end{equation}
With this formula, it is easy to see that
\begin{equation}
    g_{ab} \nabla_c \sigma^{cb} = \bar g_{ab} \bar\nabla_c \bar \sigma^{cb}.
\end{equation}
This expresses York's result \cite{York:york_method_long} that the transverse condition on the traceless part of a tensor density is conformally invariant. Collecting these results, the diffeomorphism constraints can be seen to transform as
\begin{equation}
    V_a = 2 \lf( \bar g_{ab} \bar\nabla_c \bar \sigma^{cb} + \frac 1 d \bar \nabla_a \lf( \bar \pi - \mean{\bar \pi}_{\bar g} \sqrt {\bar g} \rt) \rt) \approx 0.
\end{equation}
Using the volume preserving constraints $D$ from \eqref{eq:vpc const}, we can define an equivalent constraint surface via the modified diffeomorphism constraints
\begin{equation}
    \boxed{V_a \to 2 \bar g_{ab} \bar\nabla_c \bar \sigma^{cb} \approx 0}.
\end{equation}
These are the generators of local volume preserving diffeomorphisms.

We now collect all the first class constraints
\begin{subequations}\label{eq:York constraints}
\begin{align}
	H_\text{EY} &\equiv e^{\bar\phi_0} - \mean{e^{\frac {2d}{d- 2}\lambda_0}}_{\bar g}\approx 0 & V_a &= 2 \bar g_{ab} \bar\nabla_c \bar \sigma^{cb}\approx 0 \\
	D &= \bar \pi - \mean{\bar \pi}_{\bar g} \approx 0 & D_0 &\equiv \frac{\bar \pi_0} {V_{\bar g}} - \bar P \approx 0,
\end{align}
\end{subequations}
where we have used the fact that $\bar P = \frac 2 d \mean{\bar \pi}_{\bar g}$ and defined the \emph{extended York Hamiltonian} $H_\text{EY}$, which generates the dynamics of the entire system.

\subsection{Shape Dynamics}\label{sec:SD}

If we do not require a fixed variation of $P$, then we can attempt to perform gauge fixings of $D_0$ and $H_\text{EY}$ that reproduce different versions of ADM.

The first gauge fixing is to perform a free variation of the volume by imposing the free variation condition $\bar\pi_0 \approx 0$, freeing the global scale. This gauge fixes $H_\text{EY}$ since $\pb{H_\text{EY}}{\bar\pi_0} = e^{\bar\phi_0}$, which can not be zero for finite $\bar\phi_0$. As before, the Dirac bracket for this gauge fixing reduces to the Poisson bracket on the ADM phase space $\Gamma(\bar g_{ab},\bar \pi^{ab})$. Applying $H_\text{EY} = 0$ and $\bar\pi_0 = 0$ as strong equations, the remaining system of equations is
\begin{align}
    V_a &= 2 \bar g_{ab} \bar\nabla_c \bar \sigma^{cb}\approx 0 & D + D_0 &= \bar \pi \approx 0.
\end{align}
This is a static theory of fully conformal geometry. It is equivalent to taking the ADM theory in a gauge where $N=0$. However, this is \emph{not} a valid gauge fixing of General Relativity because $N=0$ implies that the determinant of the spacetime metric, which is given by $N\sqrt g$, is zero implying that the spacetime metric is degenerate. Consequently, the gauge fixing $\bar\pi_0 \approx 0$ is simply not valid.

There is, however, a gauge fixing that is equivalent to a valid gauge fixing of General Relativity. This is to perform a \emph{free} variation of $P$ by imposing the condition $\bar\phi_0 \approx 0$.\footnote{Note that, by adding a boundary term to the extended ADM action, we can express $\bar\pi_0$ as the configuration variable with $\bar\phi_0$ as its conjugate momentum. This justifies $\bar\phi_0\approx 0$ as the free variation condition of $P$.} This condition gauge fixes $D_0$ because $\pb{\bar\phi_0}{D_0} = V_{\bar g}^{-1}$, which is positive definite, and the Dirac bracket reduces to the Poisson bracket on the ADM phase space. The strong equations $\bar\pi_0 = 0$ and $\bar\phi_0 = 0$ lead to the system of equations
    \begin{align}
	H_\text{SD} &\equiv V_{\bar g} - \int d^dx \sqrt {\bar g} e^{\frac {2d}{d- 2}\lambda_0}\approx 0 \label{eq:SD ham}\\ V_a &= 2 \bar g_{ab} \bar\nabla_c \bar \sigma^{cb}\approx 0 \\
	D &= \bar \pi - \mean{\bar \pi}_{\bar g} \approx 0,
    \end{align}
where we have identified the Shape Dynamics Hamiltonian, $H_\text{SD}$. These are the fundamental equations of Shape Dynamics as derived in \cite{gryb:shape_dyn}. Unfortunately, this formulation of Shape Dynamics \emph{does not} implement York's ontology because it explicitly preforms a \emph{free} variation of the mean of the York time.

A final possibility is to deparametrize Shape Dynamics with respect to $\bar P$ by performing a gauge fixing of $H_\text{SD}$ using the gauge fixing condition $\bar P = t$. This gives an evolution equation for the conformal variables in terms of a specified value of the York time. Explicitly, the gauge fixing $\bar P=t$ is canonically conjugate to $H_\text{SD}$ so it can be applied strongly to eliminate $\bar P$ and $V_{\bar g}$. When these conditions are applied, the symplectic term in the canonical action containing $V_{\bar g} \dot{\bar P}$ becomes a true Hamiltonian
\begin{equation}\label{eq:dep Ham}
    H_\text{SD,deparametrized} = \int d^dx \sqrt {\bar g} e^{\frac {2d}{d- 2}\lambda_0(\bar g_{ab}, \bar \sigma^{ab}, t)}.
\end{equation}

There are at least three difficulties with this theory:
\begin{enumerate}
    \item It suffers from an `Arbitrary Choice Problem' because there is no particular reason to favour a deparametrization with respect to $\bar P$ over some other variable.
    \item Because the Lichnerowicz--York equation, $\lf. T_{\bar \lambda} S \rt|_{\bar\lambda = \lambda_0}$, depends explicitly on the York time (see \eqref{eq:LY eqn}), the deparametrized Hamiltonian \eqref{eq:dep Ham} is time \emph{dependent}. Thus, $t$ acts like a clock whose  dynamics are entangled with the system, and affect its characteristic behaviour. For this reason, the evolution does not have the key properties discussed in Section~\ref{sec:Dirty Quant} and the system does not obviously correspond to a relationally quantized system.\footnote{In the case with no cosmological constant and Higgs mass, this theory can be shown to obey a form of \emph{dynamical similarity}, as shown in \cite{Barbour:PoT_in_SD}, which may address some of these problems.}
    \item After a deparametrization w.r.t. the York time at the classical level, the York time becomes the independent parameter labelling evolution after quantization. This means that correlations between operators are compared at \emph{definite} values of this independent parameter. This has the result that the independent parameter is effectively treated as a classical object because the effect of its quantum fluctuations cannot be captured by the formalism. Singling out a particular (arbitrary) variable as being classical in this way violates relational principles and is precisely the deficiency that is cured by our Relational Quantization procedure.

\end{enumerate}
All of these problems are resolved by applying the Relational Quantization procedure discussed in the next section.

\subsection{Relational Quantization of General Relativity}\label{sec:Rel GR}

Because the ADM theory, which can be expressed in terms of the constraints \eqref{eq:York constraints}, does not permit a fixed variation of the York time, we must apply an extension procedure to the ADM theory in order to implement York's ontology. To do this, we need to add a variable to the theory that will be eliminated by solving the extended York Hamiltonian. The only way to do this and allow a fixed variation of $P$, is to remove the $\pi_0 \approx 0$ constraint from the theory. This will make dynamical exactly the mode we are looking for; since, in the transformed theory, this corresponds to removing the $D_0$ constraint, which restricts the value of the York time. The resulting theory is given by the system of constraints
\begin{subequations}\label{eq:ext SD}
    \begin{align}
     	H_\text{EY} &= e^{\phi_0} - \mean{e^{\frac {2d}{d- 2}\lambda_0}}_{\bar g} \approx 0 \label{eq:evolution} \\ V_a &= 2 \bar g_{ab} \bar\nabla_c \bar \sigma^{cb}\approx 0 \\
	D &= \bar \pi - \mean{\bar \pi}_{\bar g} \approx 0,
    \end{align}
\end{subequations}
defined on the extended phase space $\Gamma_\text{e}(\bar g_{ab}, \bar \phi_0; \bar \pi^{ab}, \bar\pi_0)$. The counting of the independently specifiable initial data goes as follows: the extended York Hamiltonian restricts the value of $(\bar\phi_0, \bar\pi_0)$, the diffeomorphism constraints restrict coordinate dependent information in the metric and momenta, and the volume preserving conformal constraints restrict the conformal factor of the metric and the trace of the momenta while leaving the volume and the York time dynamical degrees of freedom. This reproduces exactly the York ontology.

The classical evolution of this system is completely equivalent to that of classical Shape Dynamics, which, in turn, is equivalent to that of the classical ADM theory. To see this directly, we can define the convenient variables
\begin{align}
    \tau &= e^{-\bar\phi_0}\pi_0 & p_\tau &= e^{\bar\phi_0},
\end{align}
which is a canonical transformation. If we then deparametrize with respect to $\tau$ using $\tau = t$, we find that the $\bar\phi_0 \dot{\bar{\pi}}_0$ piece of the symplectic term of the canonical action reduces to a true Hamiltonian of the form
\begin{equation}
    H = \mean{e^{\frac {2d}{d-2}\lambda_0(\bar g_{ab}, \bar \sigma^{ab}, \bar P) }}_{\bar g}.
\end{equation}
This generates the same flow as the Shape Dynamics Hamiltonian \eqref{eq:SD ham} for a choice of global lapse $N = V_{\bar g}^{-1}$. There is, however, a more general way to see that our theory is classically equivalent to ADM. Note that the theory we had \emph{before} the canonical transformation was just the ADM theory with an additional global degree of freedom with no dynamics of its own. After the canonical transformation, this degree of freedom mixes with the Hamiltonian constraint and eventually becomes fixed by solving $H_\text{EY} \approx 0$. However, because we have simply performed a canonical transformation, we have not actually changed the original classical theory. The extra variable becomes an unphysical label for the evolution of the system.

The quantum theory, however, will differ because the volume and York time will become operators associated with quantum observables. This can be seen directly by noting that the quantum theory is given by Dirac quantizing the system \eqref{eq:ext SD}. In particular, the extended York Hamiltonian becomes a quantum constraint on the wavefunction
\begin{equation}
    i\hbar \diby{\ket\Psi_\text{phys}}{\tau} = \hat H \ket\Psi_\text{phys}.
\end{equation}
Because $\hat H$ is independent of $\tau$, the general solution is given by taking superpositions of eigenstates of $\hat H$
\begin{equation}
    \ket\Psi_\text{phys} = \sum_n c_n e^{iE_n \tau} \ket{\psi_n},
\end{equation}
where $\hat H \ket{\psi_n} = E_n \ket{\psi_n}$. Observables are self-adjoint operators associated with the functions on the reduced phase space obtained by gauge fixing the volume preserving conformal and diffeomorphism constraints. This is consistent with treating the global reparametrization invariance of General Relativity as a conservation symmetry. The theory we obtain has all the features of a relationally quantized theory pointed out in Section~\ref{sec:Dirty Quant}. We should not, however, understate the difficulty associated with determining the spectrum of $\hat H$, defining an inner product on the Hilbert space, and simultaneously solving the quantum volume preserving conformal and diffeomorphism constraints.

The above theory should be contrasted with the results of Wheeler--DeWitt quantization. It is easiest to compare this to the formal Dirac quantization of the ADM theory re-written in the form of Shape Dynamics. It is clear from the analysis in Section~\ref{sec:SD} that this corresponds to a \emph{free} variation with respect to the mean of the York time. Unsurprisingly, this leads, upon Dirac quantization, to a theory that selects
\begin{equation}
    \ket\Psi_\text{phys} = \ket{\psi_0}
\end{equation}
so that the zero eigenvalue is selected. This is consistent, in accordance with the general discussions in Section~\ref{sec:Dirty Quant}, with what one should expect by treating reparametrization invariance as a conventional gauge symmetry. Because the evolution is frozen, a notion of evolution must be obtained by deparametrizing with respect to a degree of freedom that one must choose arbitrarily. The observables of the theory depend on this choice and, even for simple models, can lead to complicated expressions \cite{Dittrich:2006,Dittrich:2007}. In the Relational Quantization of General Relativity, the identification of the observables is straightforward because of the time-independence of the Hamiltonian and the nature of the linear constraints. The spectrum of $\hat H$ is no more difficult to compute in the Wheeler--DeWitt case than for the case of Relational Quantization, so in this respect neither approach has an advantage. However, under Relational Quantization, the remaining constraints of the quantum theory can be interpreted as the generators of genuine gauge symmetries associated with spatial diffeomorphism and conformal invariance. Overall, this leads to a unique identification of the degrees of freedom --- not depending on an arbitrary choice of parametrization --- and is consistent with what we have identified as the characteristic behaviour of the classical theory. Such a straightforward isolation of the relevant notions of symmetry and evolution has not, to date, proved achievable within  Wheeler--DeWitt type approaches to quantum gravity. 

\section{Conclusions and Comments}

Following York, we have required that the variational principle of General Relativity should be such that the independently specifiable initial data is a conformal geometry and the mean of the York time. We then translated this `York ontology' into the precise requirement that the volume preserving part of the conformal factor be treated with a \emph{free} variation. This was shown to lead to a quantum theory with genuine evolution generated by a \emph{time-independent} Hamiltonian (given by the global constraint of \eqref{eq:ext SD}), consistent with the application of the Relational Quantization procedure discussed in Section~\ref{sec:Dirty Quant} and our previous paper \cite{gryb:2011}. The local symmetries of our theory (generated by the local constraints of \eqref{eq:ext SD}) can be classified as manifest symmetries associated with \emph{freely} varied degrees of freedom since $H_\text{EY}$ is invariant under volume preserving conformal transformations and spatial diffeomorphisms, and because we have explicitly used free variations. Thus, the local symmetries can be treated as conventional gauge symmetries. This is in contrast to the standard ADM theory where, as we saw in Section~\ref{sec:ADM}, the constraint algebra is such that a reduction of the phase space is not possible for the spatial diffeomorphisms alone unless all constraints are solved simultaneously.

A couple of further comments are in order: First, since our approach differs most significantly from Wheeler--DeWitt quantization with regard to the treatment of a single global scale (i.e., the variable conjugate to the mean of the York time), the novel features of our proposal are most evident in the setting of homogeneous cosmology. Matter can be included in our model by adding a matter Hamiltonian to the scalar ADM constraint \eqref{eq:ADM Ham}. After the canonical transformation, this will modify the solution of $T_{\lambda_0} \ham = 0$ in terms of $\lambda_0$ leading to a different Hamiltonian.\footnote{Note: this can change the uniqueness properties of these solutions \cite{MatterPaper} but not the existence of solutions in the spatially closed case.} For homogeneous cosmology, this corresponds to extra global terms in $H_\text{EY}$ that depend on the homogeneous scalar field. However, in our approach, both the conformal factor (because it is conjugate to $P$) \emph{and} the scalar field should be promoted to operators in the quantum theory. This is contrary to standard Wheeler--DeWitt cosmology where the homogeneous Wheeler--DeWitt equation (i.e., the quantum Friedmann equation) is deparametrized with respect to one of these. Thus, our model will lead to distinct predictions for early Universe cosmology. 

Second, our proposal overcomes a rather mysterious feature of the Wheeler--DeWitt equation, namely: that the Wheeler--DeWitt equation is inherently a \emph{real} equation in the sense that it does not couple real and imaginary parts of the wavefunction. It would be quite strange for such an equation to contain any limit where a complex Schr\"odinger equation \emph{does} couple real and imaginary parts of the wavefunction. In our approach, there is no mystery because our evolution equation \eqref{eq:evolution} for the wavefunction of the Universe is manifestly complex and, thus, not in conflict with the usual Schr\"odinger equation.

Finally, the role of the global scale, in particular, its connection with the evolution of the system, is a fascinating feature of our theory in need of further exploration. We have retained this variable in order to explain the characteristic behaviour of our Universe, but perhaps our analysis hints at a deeper explanation. One compelling possibility is that the global scale is linked to evolution through the \emph{double emergence} scenario explored in \cite{Barbour:Anomaly_paper}. There, a Weyl anomaly associated with the global scale of a toy model is seen to induce a behaviour similar to a renormalization group flow of the shape degrees of freedom of the theory. This behaviour is interpreted as time evolution. It may be that the flow of $H_\text{EY}$ could be described in this way. This would be similar to the Holographic cosmology scenarios discussed, for example, in \cite{Strominger:holo_cosmo,Skenderis:holo_uni}. If successful, the double emergence scenario would be a concrete realization of the Holographic principle potentially independent of String Theory considerations.

\section*{Acknowledgments}

We would like to thank Edward Anderson, Julian Barbour, Brian Pitts, Henrique Gomes, Tom Pashby, Hans Westman and  Ken Wharton for comments on the draft, and Igor Khavkine for useful discussions. SG would like to acknowledge support from and NSERC PDF grant, for travel support from Renate Loll, and for the hospitality of Utrecht and Radboud Universities. KT would like to acknowledge support of the Alexander von Humboldt foundation through a Humboldt Professorship.

\bibliographystyle{utphys}
\bibliography{mach,Masterbib}

\end{document}